\documentclass[epj,nopacs,final]{svjour}

\usepackage{graphicx}
\usepackage[switch]{lineno}
\usepackage{hyperref}
\usepackage{cite}
\usepackage{subfig}
\usepackage{multirow}
\usepackage{amsmath}
\usepackage{textcomp}
\usepackage{footnote}

\newcommand*\patchAmsMathEnvironmentForLineno[1]{%
  \expandafter\let\csname old#1\expandafter\endcsname\csname #1\endcsname
  \expandafter\let\csname oldend#1\expandafter\endcsname\csname end#1\endcsname
  \renewenvironment{#1}%
     {\linenomath\csname old#1\endcsname}%
     {\csname oldend#1\endcsname\endlinenomath}}%
\newcommand*\patchBothAmsMathEnvironmentsForLineno[1]{%
  \patchAmsMathEnvironmentForLineno{#1}%
  \patchAmsMathEnvironmentForLineno{#1*}}%
\AtBeginDocument{%
\patchBothAmsMathEnvironmentsForLineno{equation}%
\patchBothAmsMathEnvironmentsForLineno{align}%
\patchBothAmsMathEnvironmentsForLineno{flalign}%
\patchBothAmsMathEnvironmentsForLineno{alignat}%
\patchBothAmsMathEnvironmentsForLineno{gather}%
\patchBothAmsMathEnvironmentsForLineno{multline}%
}

\journalname{Eur. Phys. J. C}

\begin{document}

\title{Detailed studies of $^{100}$Mo two-neutrino double beta decay  
 in NEMO-3}



\author{
  R.~Arnold\inst{1}
  \and C.~Augier\inst{2}
  \and A.S.~Barabash\inst{3}
  \and A.~Basharina-Freshville\inst{4}
  \and S.~Blondel\inst{2}
  \and S.~Blot\inst{5}
  \and M.~Bongrand\inst{2}
  \and D.~Boursette\inst{2}
  \and V.~Brudanin\inst{6,7}
  \and J.~Busto\inst{8}
% ...
}

\institute{
  IPHC, ULP, CNRS/IN2P3, F-67037 Strasbourg, France
  \and LAL, Universit\'e Paris-Sud, CNRS/IN2P3, Universit\'e Paris-Saclay, F-91405 Orsay, France
  \and NRC ``Kurchatov Institute'', ITEP, 117218 Moscow, Russia
  \and UCL, London, WC1E 6BT, United Kingdom
  \and University of Manchester, Manchester, M13 9PL,~United Kingdom
  \and JINR, 141980 Dubna, Russia
  \and National Research Nuclear University MEPhI, 115409 Moscow, Russia
  \and Aix Marseille Universit\'e, CNRS, CPPM, F-13288 Marseille, France
% ...
}

\mail{tretyak@jinr.ru}

\date{Received: date / Accepted: date}

\abstract{
The full data set of the NEMO-3 experiment has been used to measure 
the half-life of the two-neutrino double beta decay of $^{100}$Mo
to the ground state of $^{100}$Ru,
$T_{1/2} = \left[ 6.81 \pm 0.01\,\left(\mbox{stat}\right) ^{+0.38}_{-0.40}\,\left(\mbox{syst}\right) \right] \times10^{18}$~y.
The two-electron energy sum, single electron energy spectra and distribution
of the angle between the electrons
are presented with an unprecedented statistics of $5\times10^5$ events and a 
signal-to-background ratio of $\sim$80.
Clear evidence for the Single State Dominance model is found for 
this nuclear transition. 
Limits on Majoron emitting 
neutrinoless double beta decay modes with spectral indices of n=2,3,7,
 as well as constraints on Lorentz invariance violation and 
on the bosonic neutrino contribution to the two-neutrino double beta decay mode are obtained.
}

\maketitle





\section{Introduction}
\label{Introduction}

Spontaneous nuclear double beta decay is a second order 
weak interaction process that was theoretically considered
for the first time by M.~Goeppert-Mayer in 1935~\cite{Goeppert-Mayer}.
It can occur in some even-even nuclei when two bound neutrons simultaneously undergo beta
decay and are transformed into two bound protons emitting two electrons and two (anti)neutrinos.
Two-neutrino double beta decay, $2\nu\beta\beta$, is one of the rarest directly observed radioactive processes
with half-lives ranging from $7\times10^{18}$ to $2\times10^{21}$ years ~\cite{PDG,Barabash-2nu}.

The decay rate of $2\nu\beta\beta$ decay can be expressed as 
\begin{equation}
1/T^{2\nu}_{1/2} =g_{A}^4 G^{2\nu} |M^{2\nu}|^2~, 
\label{eq:one}
\end{equation}
where $g_{A}$ is the axial-vector coupling constant, $G^{2\nu}$ is a phase space factor,
and $M^{2\nu}$ is a nuclear matrix element (NME). Measurement of the $2\nu\beta\beta$
half-life gives direct access to the value of the NME for this process and therefore 
provides experimental input into 
nuclear models that are used to evaluate NMEs. Moreover, $2\nu\beta\beta$ may provide 
answers to the question of $g_{A}$ quenching in nuclear matter 
that is currently being actively discussed \cite{Barea-2015,Pirinen-2015,Kostensalo-2017,Suhonen-2017}.
Detailed studies of $2\nu\beta\beta$ may therefore be useful to improve NME calculations for 
the neutrinoless mode 
of double beta decay, $0\nu\beta\beta$,  
the process which violates total lepton number and is one 
of the most sensitive probes of physics beyond the Standard Model. 
A recent review of the $0\nu\beta\beta$ NME calculation methods,  challenges and prospects  
can be found in~\cite{review-0nu}.

Previous measurements have  shown that the $^{100}$Mo $2\nu\beta\beta$ half-life is shorter compared 
to other  $\beta\beta$ isotopes \cite{Elegant,Nemo2,Moe,SiLi,Ashitkov,Mo100_2005,Bolometr,Armengaud-2017},
and it is therefore a promising nucleus for precise studies of the process. 
Here we present the most accurate to date study of $^{100}$Mo $2\nu\beta\beta$ decay 
including single electron energy and angular distributions of the electrons emitted in the decay with 
an unprecedented statistics of $5\times10^5$ events. The impact of the single electron energy spectra 
on nuclear models that are used to calculate the  NME is also presented. 

Searches for most commonly discussed $0\nu\beta\beta$ mechanisms (exchange of a light Majorana
neutrino, right-handed currents, super-symmetry) with NEMO-3 have been reported earlier in~\cite{0nu-short,0nu-PR}.
In this paper we present results obtained for $^{100}$Mo $0\nu\beta\beta$ decay accompanied by the emission of 
Majoron bosons with spectral indices $n\ge2$, as well as constraints on 
contributions from bosonic neutrinos and from
Lorentz invariance violation to $2\nu\beta\beta$ spectra of $^{100}$Mo. 

\section{The NEMO-3 detector}
\label{detector}
The NEMO-3 detector, its calibration and performance are described in detail in~\cite{TDR}
and more recently in~\cite{0nu-PR}.
A combination of tracking and calorimetric approaches allows 
for a full reconstruction of  $\beta\beta$ event topology. A tracking 
chamber is used to reconstruct electron tracks, their origin and end points. 
The electron energies and arrival times are measured with a plastic scintillator calorimeter.  
The cylindrical detector measuring 3~m in height and 5~m in diameter is 
made up of 20 wedge-shaped sectors of identical size. 
Each sector hosts 7 thin foil strips containing a $\beta\beta$ isotope. 
The source foils are positioned in the middle of the tracking detector at a 
radius of 1~m and have a height of 2.48~m. 

The tracking detector is based on a wire chamber made of 
6180 open drift cells operating in Geiger mode with helium as 
the main working gas with the addition of ethanol (4\%), argon (1\%) and 
water vapour (0.15\%). The wire cells are strung vertically parallel to the source foils 
and have average transverse and longitudinal resolutions of 0.5 mm
and 0.8 cm ($\sigma$) respectively. The tracking volume is
surrounded by a segmented calorimeter composed of 1940 optical modules
made of 10~cm thick polystyrene scintillator blocks coupled to 
low radioactivity photomultiplier tubes (PMT). The energy resolution 
of optical modules for 1 MeV electrons ranges from 5.8\% to 7.2\% and the time resolution is 250~ps ($\sigma$).
The detector was calibrated by deploying $^{207}$Bi, $^{90}$Sr and $^{232}$U sources during the course of data collection.
The stability of the PMT gains was monitored by 
a dedicated light injection system that was run every 12 hours. 

The NEMO-3 detector is supplied with a solenoid which generates a 25\,G magnetic field 
parallel to the tracking detector wires and provides charge identification by track curvature. 
The detector is surrounded by passive shielding consisting of a 19\,cm thick iron plates 
to suppress the external gamma ray flux, and of borated water,
paraffin and wood to moderate and absorb environmental neutrons.

One of the unique advantages of the NEMO-3 technology is the ability to unambiguously identify 
electrons, positrons, gamma- and delayed alpha-particles. This approach leads to a strong suppression of 
backgrounds by eliminating events that do not exhibit a $\beta\beta$ topology. 
In addition, it allows for an efficient background evaluation by selecting event topologies 
corresponding to specific background channels. 
An electron is identified by a reconstructed prompt track in the drift chamber
matching to a calorimeter deposit. 
Extrapolating the track to the foil plane defines the event vertex in the source. 
The track extrapolation to the calorimeter identifies the impact point of the electron track 
with the corresponding optical module and is used to correct the reconstructed energy 
of the electron deposited in the scintillator. The track curvature in the magnetic field is used
to distinguish electrons from positrons.
A $\gamma$-ray is identified as an energy deposit in the calorimeter 
without an associated track in the drift chamber. An $\alpha$-particle 
is identified by a short straight track delayed with respect to the prompt electron 
in order to tag $^{214}$Bi $\rightarrow$ $^{214}$Po delayed coincidences. 

The NEMO-3 detector took data at the Modane Underground Laboratory (LSM)
in the Frejus tunnel at a depth of 4800~m w.e. enabling the cosmic muon flux suppression by a 
factor of $>$ 10$^{6}$. 
The detector hosted source foils of 7 different $\beta\beta$ isotopes. The two isotopes 
with the largest mass were $^{100}$Mo (6.914 kg)~\cite{0nu-PR} and $^{82}$Se 
(0.932 kg)~\cite{Se82} with smaller amounts of $^{48}$Ca,  $^{96}$Zr, $^{116}$Cd, $^{130}$Te 
and $^{150}$Nd~\cite{Ca48,Zr96,Cd116,Te130,Nd150}.

Two types of purified molybdenum foils were installed in NEMO-3, metallic and composite. 
Both foil types were enriched in  $^{100}$Mo with the isotopic enrichment factor ranging from 
$95.14 \pm 0.05\%$ to $98.95 \pm 0.05\%$. The average enrichment factor was 
97.7\% for metallic foils and 96.5\% for composite foils.
The metallic foils contained $2479\pm 5$~g of $^{100}$Mo.
The mean metallic foil density is 58~mg/cm$^2$ with a total foil surface of 43924~cm$^2$.
The composite foils contained $4435\pm 13$~g of $^{100}$Mo.
They were produced by mixing a fine molybdenum powder with polyvinyl alcohol (PVA) glue 
and deposited between Mylar foils of 19 $\mu$m thickness. 
The average surface density of the composite foils is 66~mg/cm$^2$ and the total 
foil surface area is 84410~cm$^2$.

Monte Carlo (MC) simulations are performed with a GEANT-3 based ~\cite{geant3} program using the
DECAY0~\cite{decay0} event generator. The time-dependent status and performance of 
the detector are taken into account in modelling the detector response.

The data presented here were collected between February 2003 and October 2010
with a live time of 4.96 y and a total exposure of 34.3~kg$\cdot$y of $^{100}$Mo.
This is the same exposure as that used for $0\nu\beta\beta$ results published 
earlier~\cite{0nu-PR}.

\section{Background model}
\label{bkg}

Trace quantities of naturally-occurring radioactive isotopes can occasionally produce 
two-electron events and thus can mimic $\beta\beta$-decay events. 
The largest contributions come from isotopes that are 
progenies of $^{238}$U ($^{234m}$Pa, $^{214}$Pb, $^{214}$Bi, $^{210}$Bi) 
and of $^{232}$Th ($^{228}$Ac, $^{212}$Bi, $^{208}$Tl), as well as $^{40}$K.

\begin{figure}
\subfloat[Internal]{\frame{\includegraphics[width=0.48\textwidth]{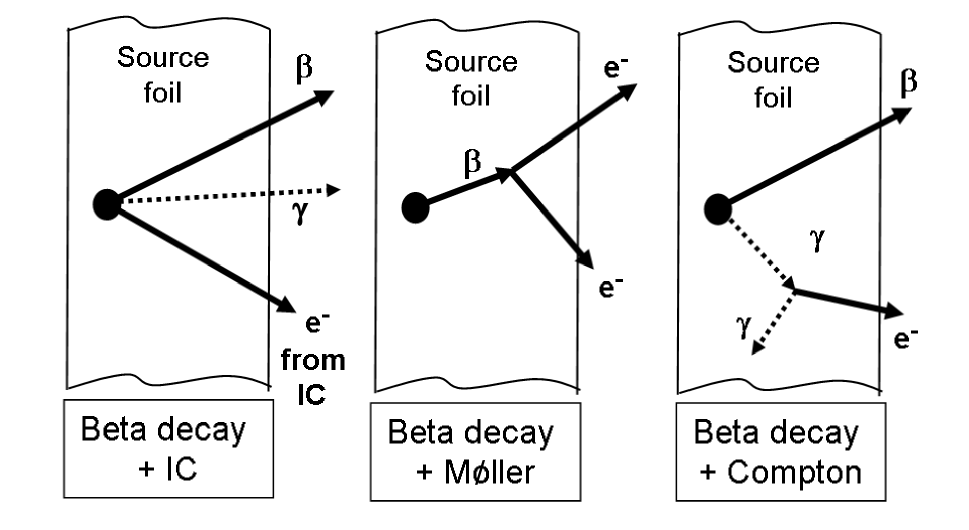}}}\\
\subfloat[External]{\frame{\includegraphics[width=0.48\textwidth]{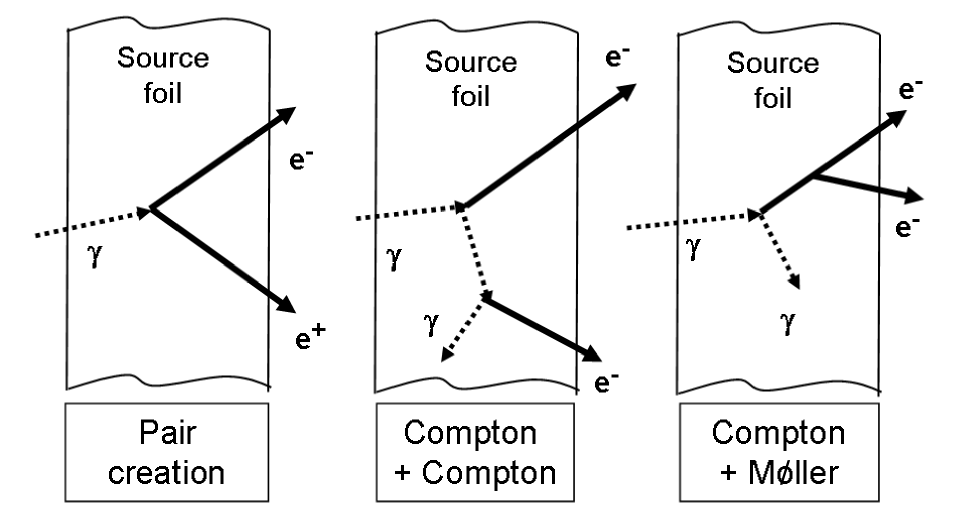}}}
\caption{Mechanisms of internal (a) and external (b) background production in the source foil}
\label{fig:inbg}
\end{figure}

The background is categorised as internal if it originates from radioactive decays 
inside  the $\beta\beta$ source foils, see Fig.~\ref{fig:inbg}(a). 
Two electrons can be produced via $\beta$-decay followed by a M{\o}ller scattering, 
$\beta$-decay to an excited state with the subsequent internal conversion or due to 
Compton scattering of the de-excitation photon. 
Decays inside the tracking detector volume form a separate background category. 
The main source of this background is radon, $^{222}$Rn. The decay of radon progenies near the
source foil can produce signal-like events in an analogous
manner to internal background decays.

The last background category is due to the external $\gamma$-ray flux 
produced by decay of radioactive isotopes in detector components, the surrounding area 
and due to neutron interactions in the shield and material of the detector.
The PMT glass is the main source of these $\gamma$-rays. They can produce 
two-electron events due to $e^+ e^-$ pair creation in the source foil and subsequent 
charge misidentification, double Compton scattering 
or Compton scattering followed by M{\o}ller scattering, see Fig.~\ref{fig:inbg}(b).

A detailed discussion of the NEMO-3 background model is presented in~\cite{bdf}
and results of screening measurements can be found in~\cite{0nu-PR,TDR,bdf}.
Here we follow the same background model as that presented for the $^{100}$Mo 
$0\nu\beta\beta$ analysis~\cite{0nu-PR}. However, radioactive isotopes contributing 
to the low energy region of the $^{100}$Mo $2\nu\beta\beta$ spectrum were not relevant 
for the $0\nu\beta\beta$ analysis in~\cite{0nu-PR} and are therefore discussed in more detail below. 
The background in question comes from traces of $\beta$-decaying isotopes 
$^{210}$Bi, $^{40}$K and $^{234m}$Pa in $^{100}$Mo foils. 
In addition, $^{100}$Mo $2\nu\beta\beta$ decay to the $0^+_1$ excited state of $^{100}$Ru
is also taken into account as a source of internal background. The experimental  half-life value
of $T_{1/2} = 6.7^{+0.5}_{-0.4} \times 10^{20}$~y~\cite{Barabash-2nu}
is used to evaluate this contribution.

\begin{figure*}
\begin{center}
\includegraphics[width=0.48\textwidth]{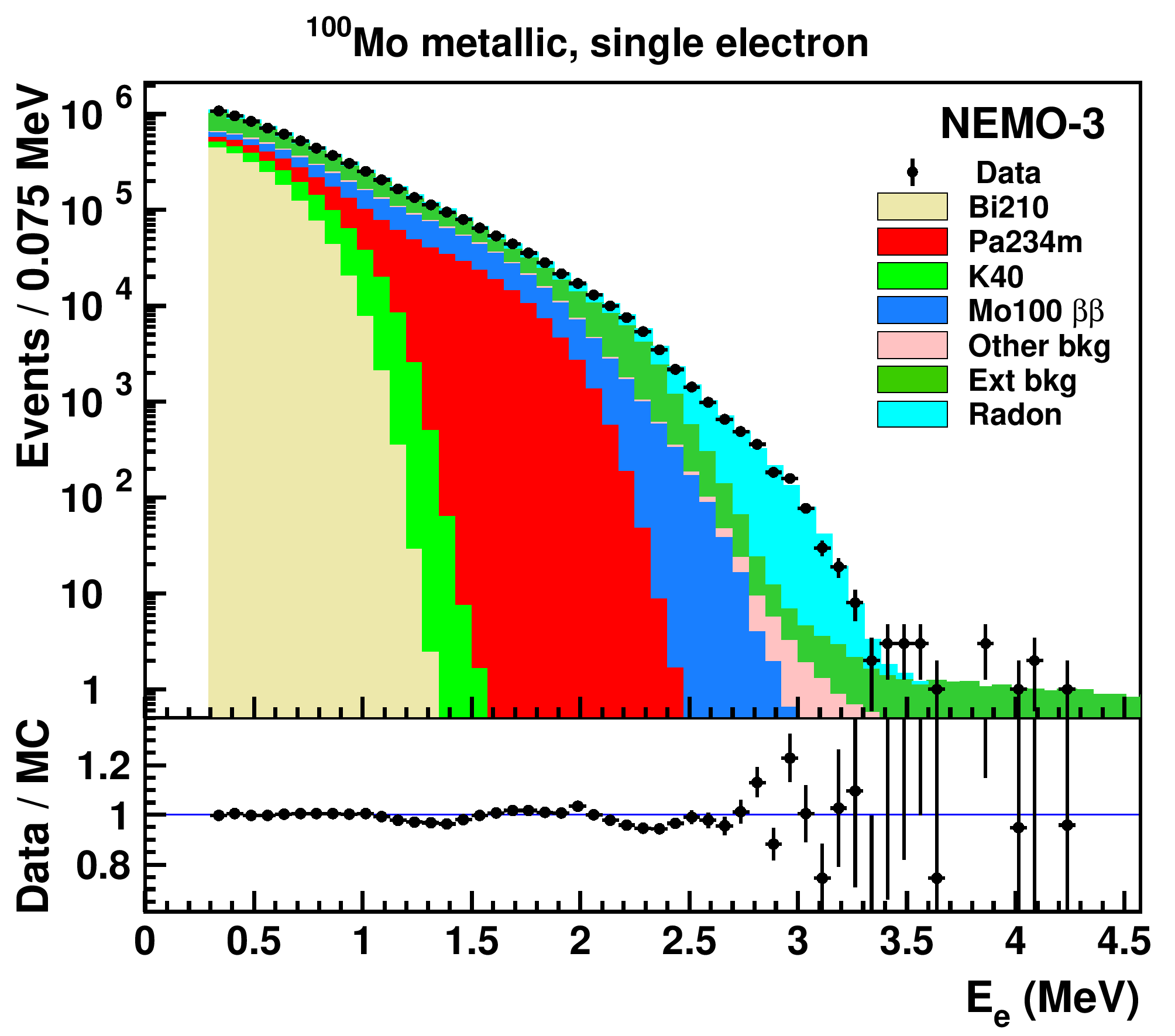}
~~~~~
\includegraphics[width=0.48\textwidth]{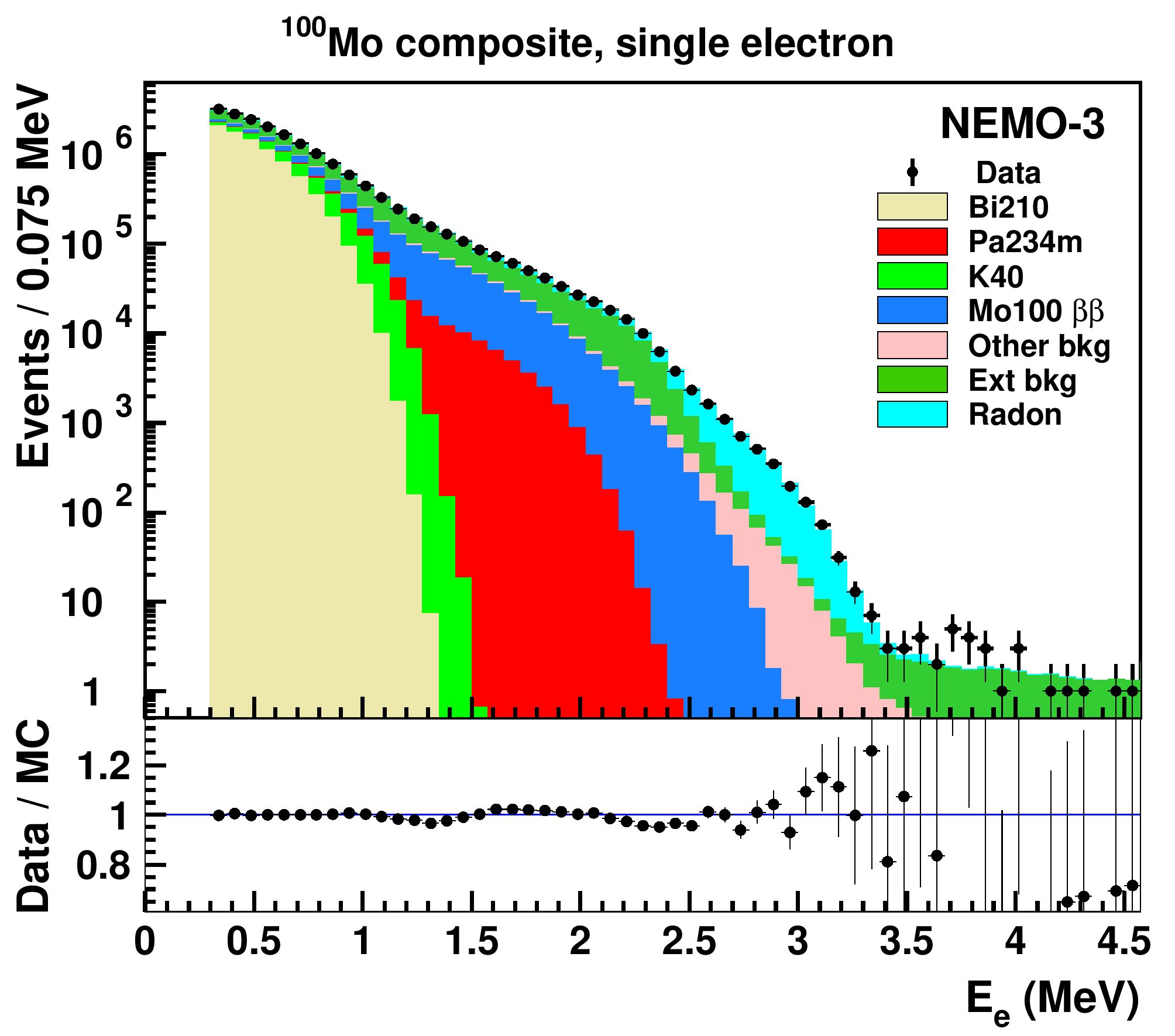}
\caption{Single electron events energy spectra for metallic and composite
molybdenum. The error bars correspond to statistical uncertainty only.
}
\label{fig:1e}
\end{center}
\end{figure*}

The activities of $\beta$-emitters in $^{100}$Mo foils are determined from the 
fit to the electron energy distribution for a single electron event sample, which 
is shown in Fig.~\ref{fig:1e} separately for metallic and composite foils. 
To disentangle the $^{210}$Bi contribution from the source foils and the surface of the tracker wires 
the activity measured in~\cite{bdf} is used for the latter. Fig.~\ref{fig:1e} shows 
the sum of both contributions. Secular equilibrium is assumed between $^{214}$Pb and $^{214}$Bi.
The same is done between  $^{228}$Ac, $^{212}$Bi and $^{208}$Tl, where the
branching ratio of 35.94\% is taken into account.
There is sufficiently good agreement between data and MC for the 
single electron energy spectrum. The observed deviations of MC from data
are within 6\% and are not significant when the systematic uncertainty 
on the external background is taken into account.

The results of the internal $^{100}$Mo foil contamination measurements carried 
out with the NEMO-3 detector are shown in Table~\ref{table:inbg}.

\begin{table}[hbt]
\caption{\label{table:inbg}%
$^{100}$Mo source foil contamination activities 
measured with the NEMO-3 detector. Activities of $^{214}$Bi and $^{208}$Tl
are from ~\cite{bdf}. }
\begin{tabular*}{\columnwidth}{@{\extracolsep{\fill}}lcc@{}}
 \hline\noalign{\smallskip}
  Source & $^{100}$Mo metallic & $^{100}$Mo composite \\
 \noalign{\smallskip}\hline\noalign{\smallskip}
$^{214}$Bi internal, mBq/kg& $0.060\pm0.019$ & $0.305\pm0.038$ \\
$^{214}$Bi mylar, mBq/kg      & $ - $           & $1.05\pm0.06$  \\
$^{208}$Tl, mBq/kg            & $0.087\pm0.004$ & $0.128\pm0.003$ \\
 \noalign{\smallskip}\hline\noalign{\smallskip}
$^{234m}$Pa, mBq/kg           & $11.40\pm0.06$    & $2.10\pm0.03$ \\
$^{40}$K , mBq/kg             & $8.67\pm0.05$    & $13.57\pm0.04$ \\
 \noalign{\smallskip}\hline\noalign{\smallskip}
$^{210}$Bi, mBq/m$^2$ & $5.51\pm0.03$   & $19.42\pm0.03$ \\
  \noalign{\smallskip}\hline
\end{tabular*}
\end{table} 
\section{Two-neutrino double beta decay of $^{100}$Mo}
\label{sec:2vbb}

Candidate $\beta\beta$ events are selected by requiring two reconstructed electron tracks, each 
associated with an energy deposited in an individual optical module. 
The energy deposited by the electron in a single optical module should be greater than 300~keV. 
Each PMT must be flagged as stable according to the light injection survey~\cite{0nu-PR}. 
The tracks must both originate from the $^{100}$Mo  source foil, and their
points of intersection with the plane of the source foil must be
within 4cm transverse to and 8cm along the direction of the tracker wires,
in order to ensure that the two tracks are associated to a common event
vertex.
The track curvatures must be consistent with electrons moving outwards 
from the source foil.
The timing and  the path length of the electrons must be 
consistent with the hypothesis of simultaneous
emission of two electrons from a common vertex in the $^{100}$Mo source 
foil~\cite{0nu-PR}.
There should be no $\gamma$-ray hits and $\alpha$-particle tracks in the event.

After the above event selection there are 501534 $^{100}$Mo two-electron candidate events, 
with 193699 coming from the metallic foils and 307835 from the composite foils. 
Table~\ref{table:bb_bkg} shows the number of expected background 
and candidate signal events in $^{100}$Mo foils. 
The number of $2\nu\beta\beta$ events is obtained from a binned log-likelihood 
fit to the two-electron energy sum distribution
under the Single State Dominance (SSD) nuclear model, as detailed below. 
The average  signal-to-background ratio is S/B=79, with S/B=63 for the metallic 
foils and S/B=94 for the composite foils. 
The detector acceptance and selection efficiency for $2\nu\beta\beta$ $^{100}$Mo events 
calculated using MC simulations is $\epsilon = (2.356\pm0.002)\%$, with 
$\epsilon_{met} = (2.472\pm0.003)\%$ and $\epsilon_{com} = (2.292\pm0.002)\%$ 
for the metallic and composite molybdenum foils respectively. 
Using the above values gives the $^{100}$Mo $2\nu\beta\beta$-decay half-life of 
$T_{1/2} = (6.65 \pm 0.02 ) \times10^{18}$~y for the metallic foils 
and $T_{1/2} = (6.91 \pm 0.01 ) \times10^{18}$~y for the composite foils. 
The difference between the two sample measurements may be 
explained by inaccuracy of the thin foil modelling 
and is taken into account in estimation of the systematic uncertainty in Section~\ref{sec:syst}.
We consider the mean value over the two data samples
as the more reliable half-life estimation
\begin{equation}
T_{1/2} = (6.81 \pm 0.01 ) \times10^{18}~\mbox{y}.
\end{equation}

\begin{figure*}[p]
\begin{center}
\includegraphics[width=0.45\textwidth]{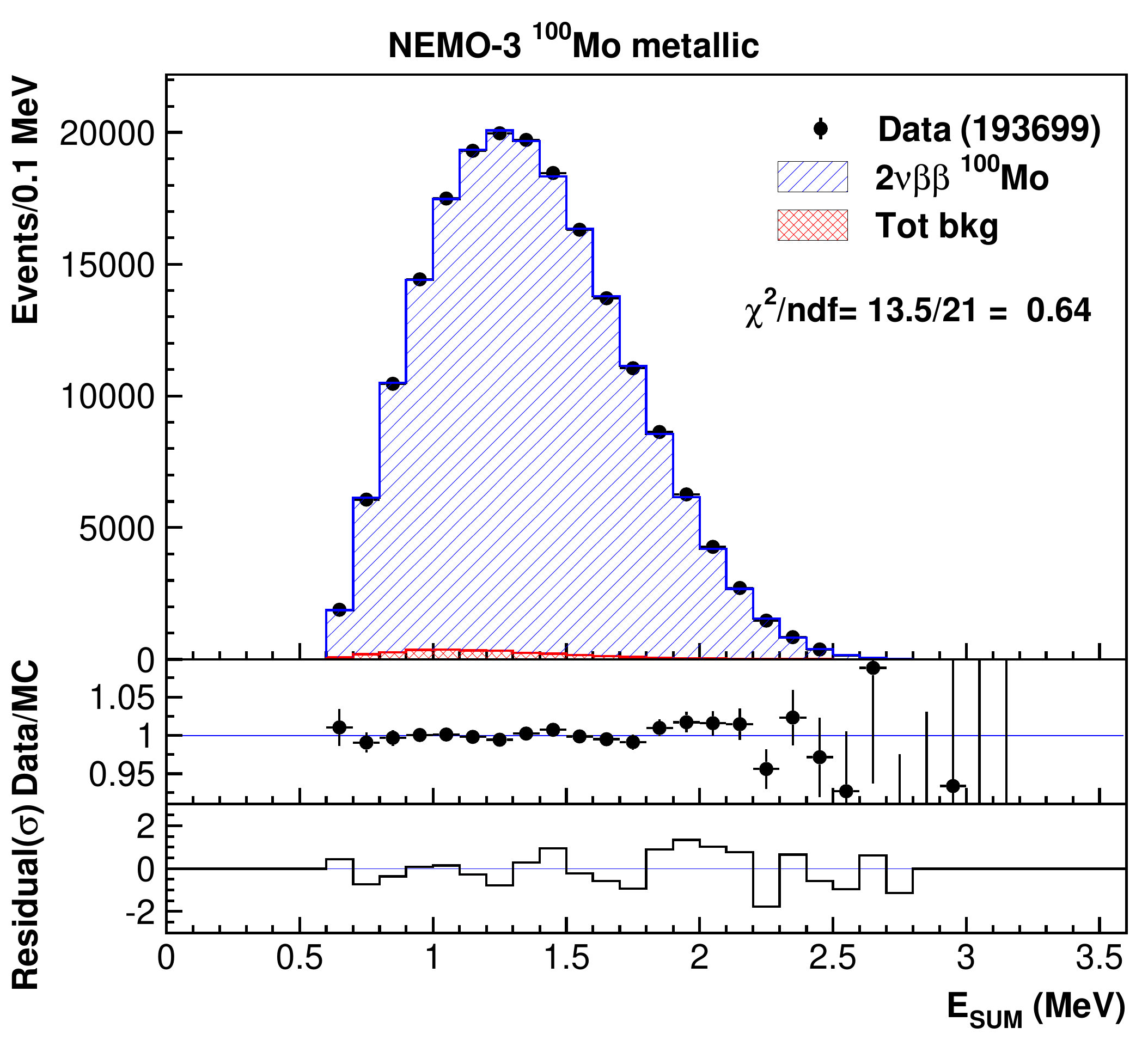}
\includegraphics[width=0.45\textwidth]{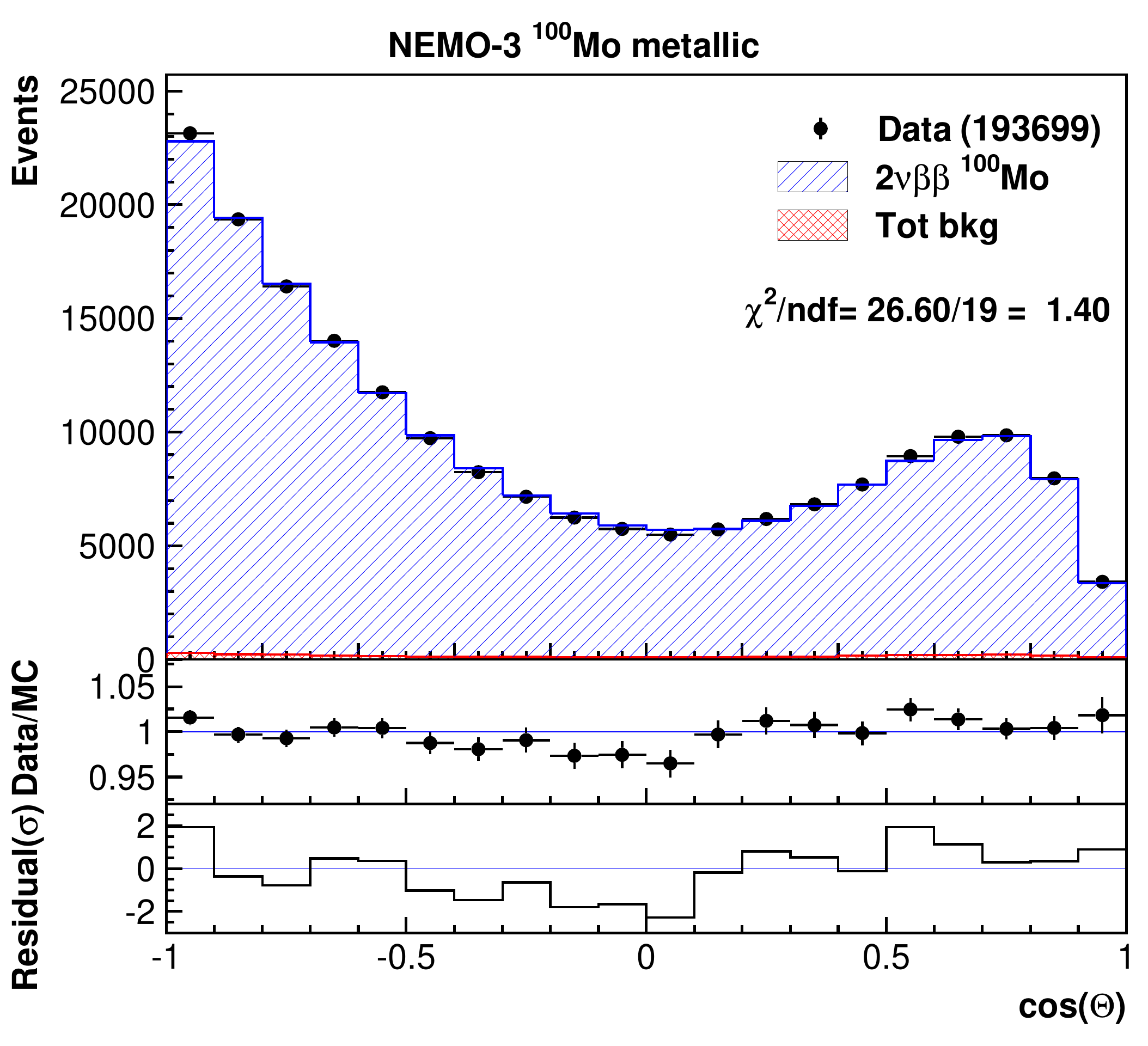}
\includegraphics[width=0.45\textwidth]{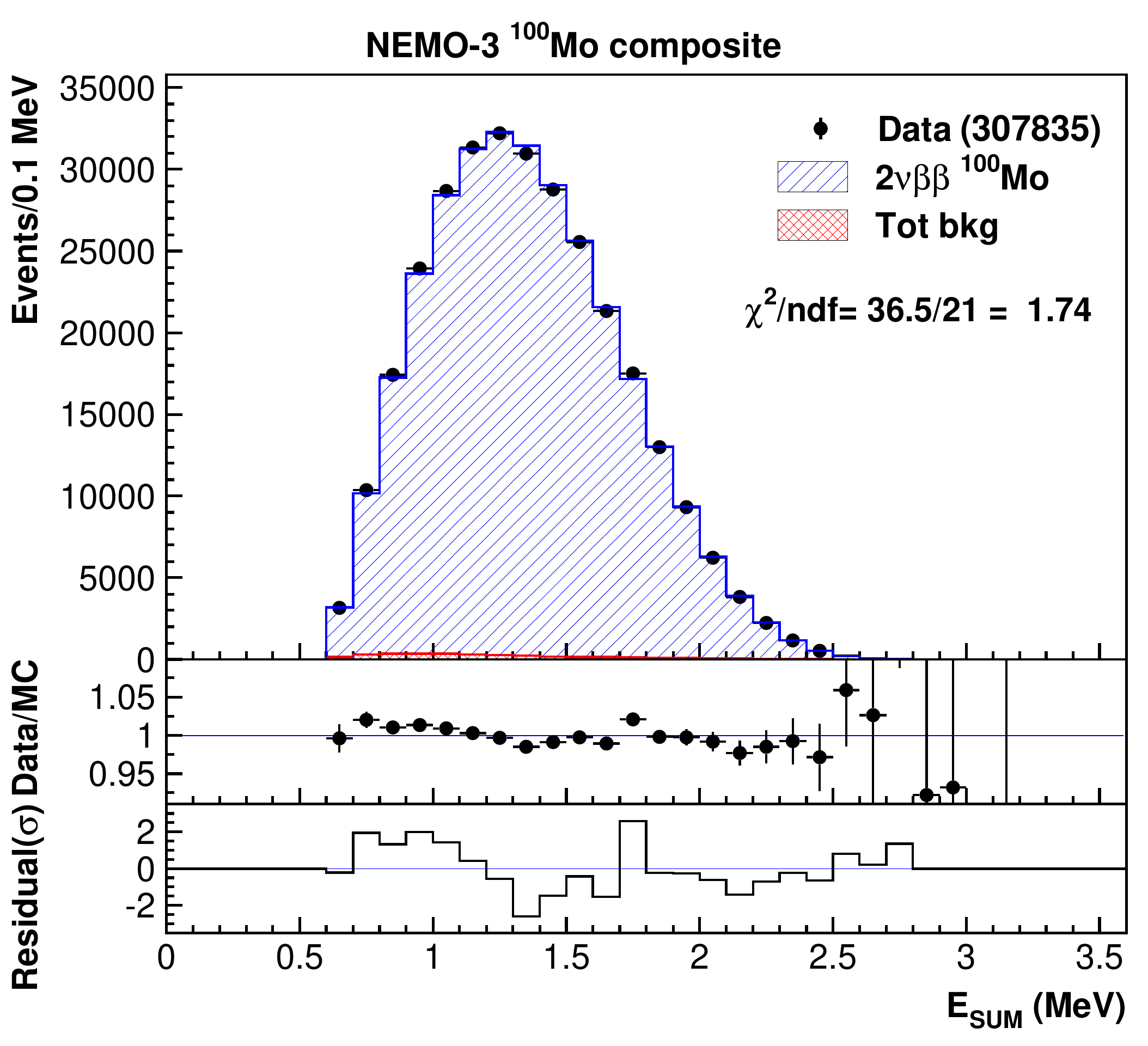}
\includegraphics[width=0.45\textwidth]{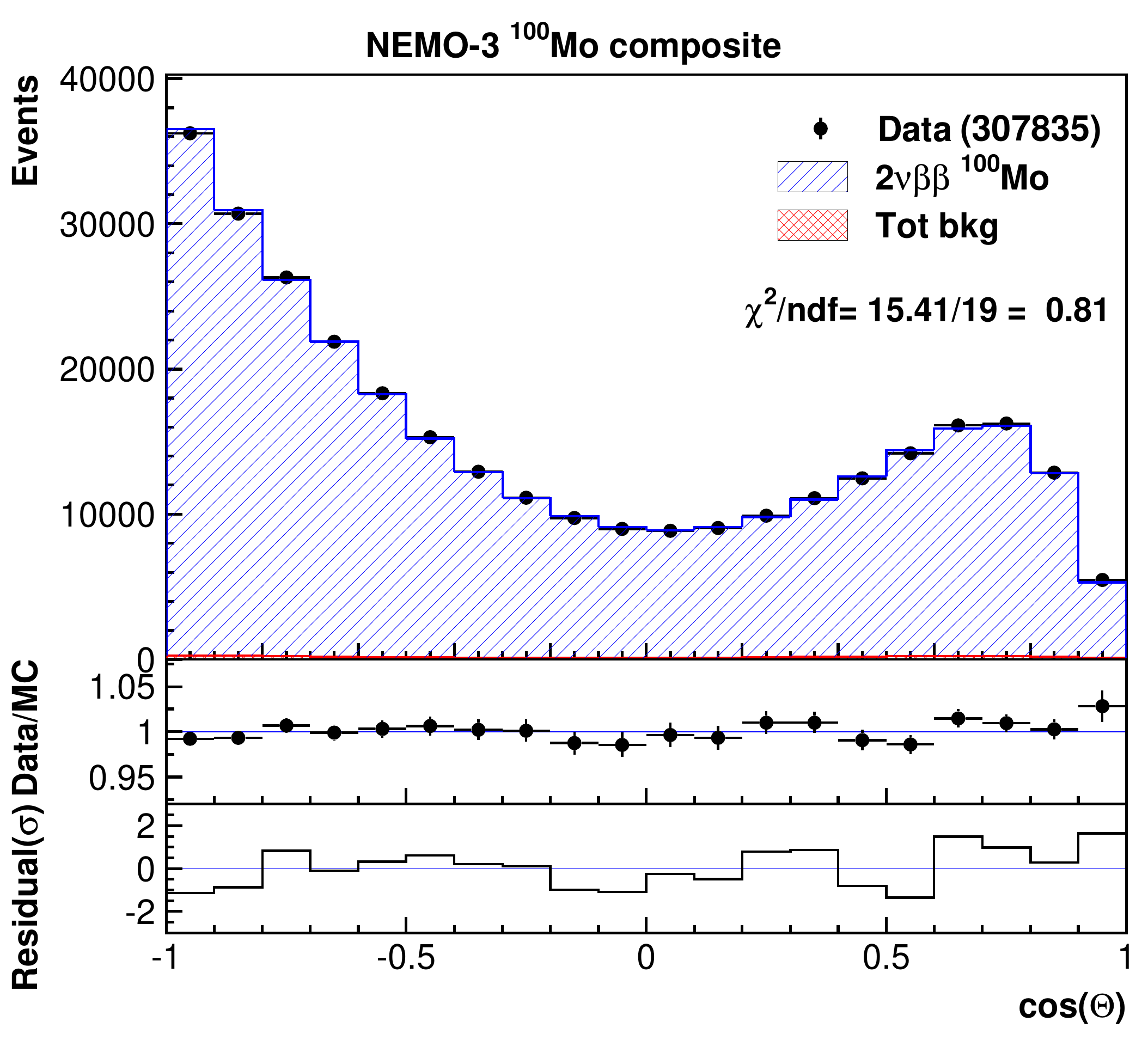}
\includegraphics[width=0.45\textwidth]{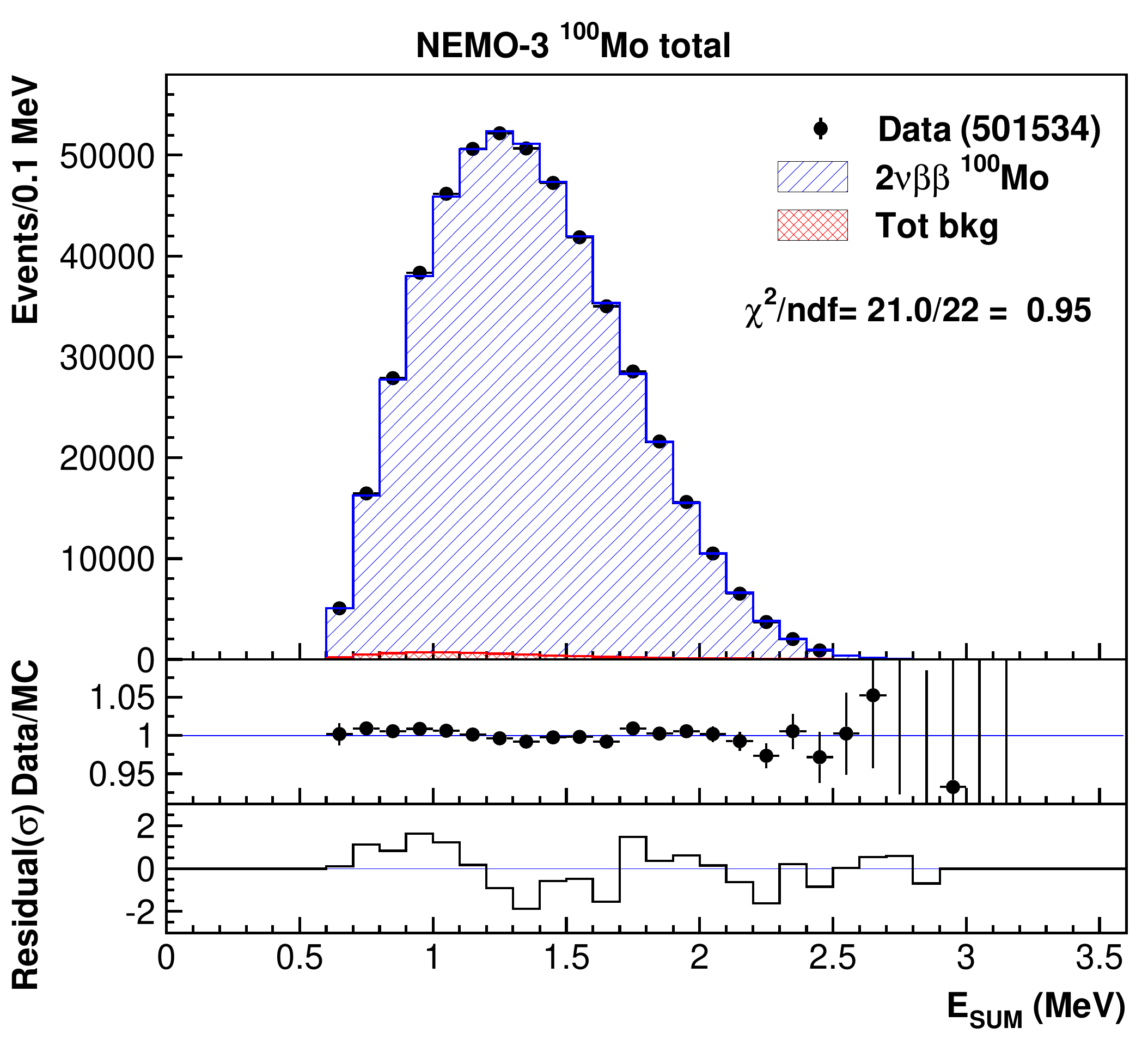}
\includegraphics[width=0.45\textwidth]{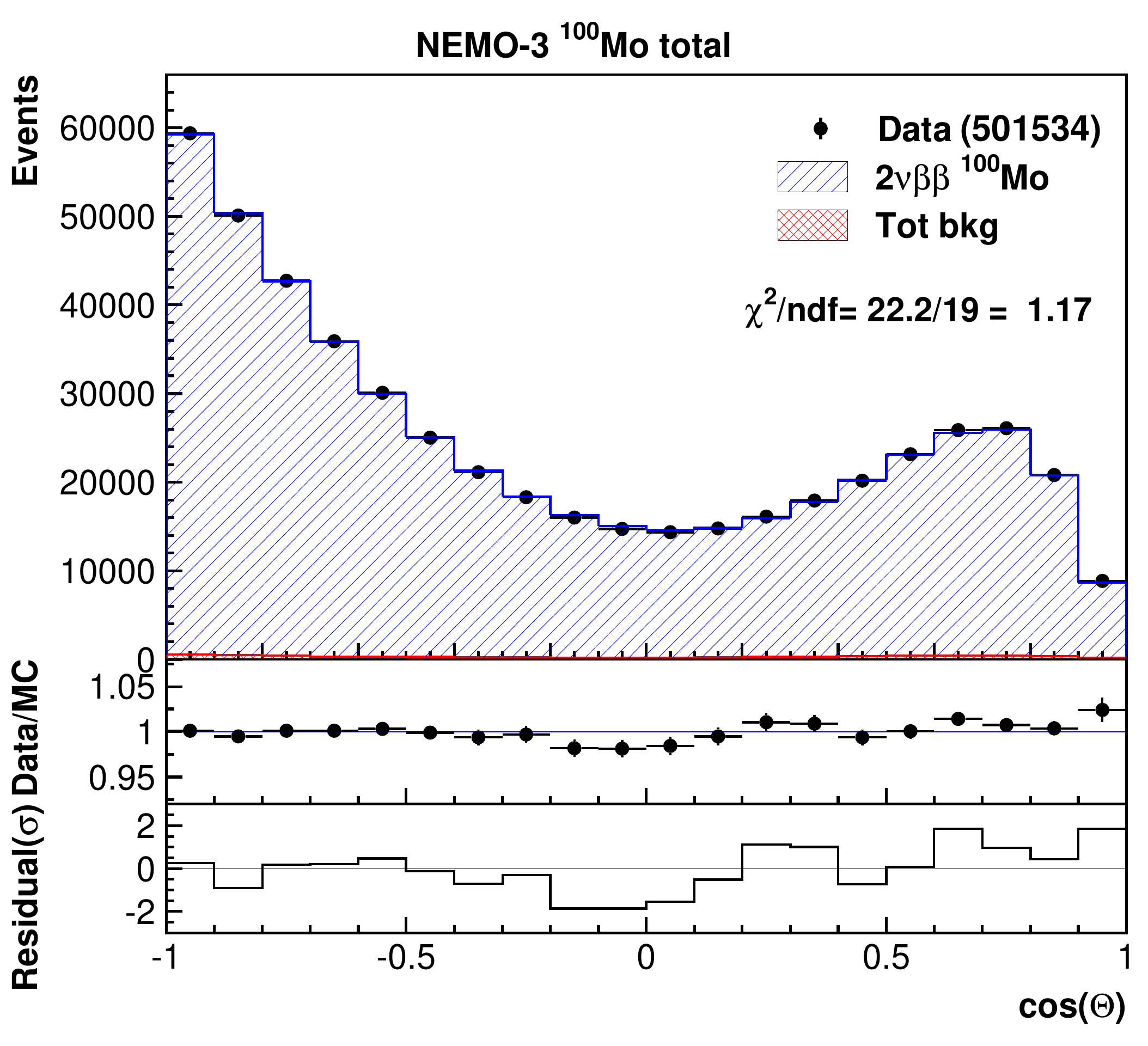}
\caption{Distributions of two-electron summed kinetic energy and the opening angle 
between two electron tracks in $^{100}$Mo foils
after an exposure of 34.3~kg$\cdot$y. Data are compared to the MC prediction 
of the SSD model (see text), where the resulting event numbers are taken
from a binned log-likelihood fit.}
\label{fig:bb_energy}
\end{center}
\end{figure*}

\begin{table}[hbt]
\caption{
Expected number of background events in the two-electron
channel and the number of $^{100}$Mo $2\nu\beta\beta$ candidate events in molybdenum foils. 
}
\begin{tabular*}{\columnwidth}{lccc}
\hline\noalign{\smallskip}
  Source    &   Metallic    &    Composite     &  Total $^{100}$Mo  \\
\noalign{\smallskip}\hline\noalign{\smallskip}
  $^{228}$Ac,$^{212}$Bi,\\  
  $^{208}$Tl       & $  49.5 \pm    0.5$ & $    142.3 \pm    1.3$ & $    191.8 \pm    1.4$\\  
  $^{214}$Pb,$^{214}$Bi        & $   14.2 \pm    0.1$ & $    177.2 \pm    0.7$ & $    191.3 \pm    0.7$\\  
  $^{40}$K          & $  101.4 \pm    2.5$ & $    296.0 \pm    7.3$ & $    397.5 \pm    7.7$\\  
  $^{234m}$Pa       & $ 1783.8 \pm   11.8$ & $    656.7 \pm    4.3$ & $   2440.5 \pm   12.5$\\  
  $^{210}$Bi        & $   25.6 \pm    1.4$ & $     90.3 \pm    2.8$ & $    115.9 \pm    3.1$\\  
  Radon             & $  434.3 \pm    6.2$ & $    590.3 \pm    5.2$ & $   1024.6 \pm    8.1$\\  
  Ext Bkg           & $  562.7 \pm    9.7$ & $   1238.6 \pm   14.7$ & $   1801.3 \pm   17.6$\\  
$\beta\beta$ $0^+_1$& $   48.6 \pm    0.8$ & $     71.1 \pm    1.0$ & $    119.7 \pm    1.3$\\  
\noalign{\smallskip}\hline\noalign{\smallskip}
  Tot bkg           & $   3020 \pm   17$   & $   3263 \pm   18$     & $   6283 \pm   25$\\  
$\beta\beta$ g.s.   & $ 190683 \pm  117$   & $ 304571 \pm  144$     & $ 495254 \pm  186$\\  
  Data              & $ 193699           $ & $    307835          $ & $   501534           $\\  
\noalign{\smallskip}\hline
\end{tabular*}
\label{table:bb_bkg}
\end{table} 
The two-electron energy sum spectra and the distributions of cosine of the angle between two electrons 
emitted from $^{100}$Mo foil are shown in Fig.~\ref{fig:bb_energy}, separately 
for the metallic and composite foils  as well as for the total $^{100}$Mo sample. 

The electron energy measured in the calorimeter is smaller 
than the energy at the point of origin due to energy losses in the foil 
and in the drift chamber. 
For instance in the case of $^{100}$Mo $2\nu\beta\beta$ decay the mean electron track length 
from the source foil to the calorimeter is 75~cm and the mean energy loss of electrons in the drift chamber is 43~keV.
The single and summed electron energy distributions are
presented for the measured values of the electron kinetic energy
$E_e$ and sum of the measured electron kinetic energies $E_{SUM}$, 
respectively, i.e., without correction for the energy loss.

The angular distribution is corrected with the well-measured distribution of the opening angle between two electrons emitted in $^{207}$Bi decay. The MC distribution of the cosine of the angle between two electron tracks has been reweighted based on data collected in the regular energy calibration runs performed with $^{207}$Bi sources. The correction is biggest for small opening angles, and is at the level of 4\% on average.

\subsection{Role of intermediate nuclear states in $^{100}$Mo $2\nu\beta\beta$ transition}
The nuclear $\beta\beta$ decay (A,Z) $\to$ (A,Z+2) is realized via two subsequent virtual 
$\beta$ transitions through the complete set of states
of intermediate nucleus (A,Z+1).
In the case of $^{100}$Mo $2\nu\beta\beta$ transition 
between the ground states of the parent ($^{100}$Mo) and daughter ($^{100}$Ru) 
nuclei with spin-parity $0^+$
the process is governed by two Gamow-Teller transitions through
$1^+$ states of $^{100}$Tc.
Nuclear theory does not predict a priori 
whether there is a dominance of transition through
the $1^+$ ground state (SSD hypothesis~\cite{ssd_abad,ssd_simkovic}) or through higher
lying excited states, namely from the region of the Gamow-Teller
resonance (HSD hypothesis). The SSD versus HSD analysis is feasible
as the ground state of $^{100}$Tc has spin-parity $J^P=1^+$
and is lying close to the ground state of $^{100}$Mo.

The evidence in favour of SSD in $^{100}$Mo $2\nu\beta\beta$ decay
was already observed at the beginning of NEMO-3 data analysis~\cite{ssd_shitov}.
Further hints for the SSD model in the $^{100}$Mo $2\nu\beta\beta$ decay were obtained in charge-exchange 
experiments by observing a strong Gamow-Teller transition to the $1^+$ ground state of $^{100}$Tc 
in the $^{100}$Mo($^{3}$He,t)$^{100}$Tc reaction~\cite{ssd_thies}. It was estimated 
that this transition could contribute 
as much as 80\% to the total value of the $^{100}$Mo $2\nu\beta\beta$ matrix element. 

\begin{figure}[htb]
\begin{center}
\includegraphics[width=0.35\textwidth]{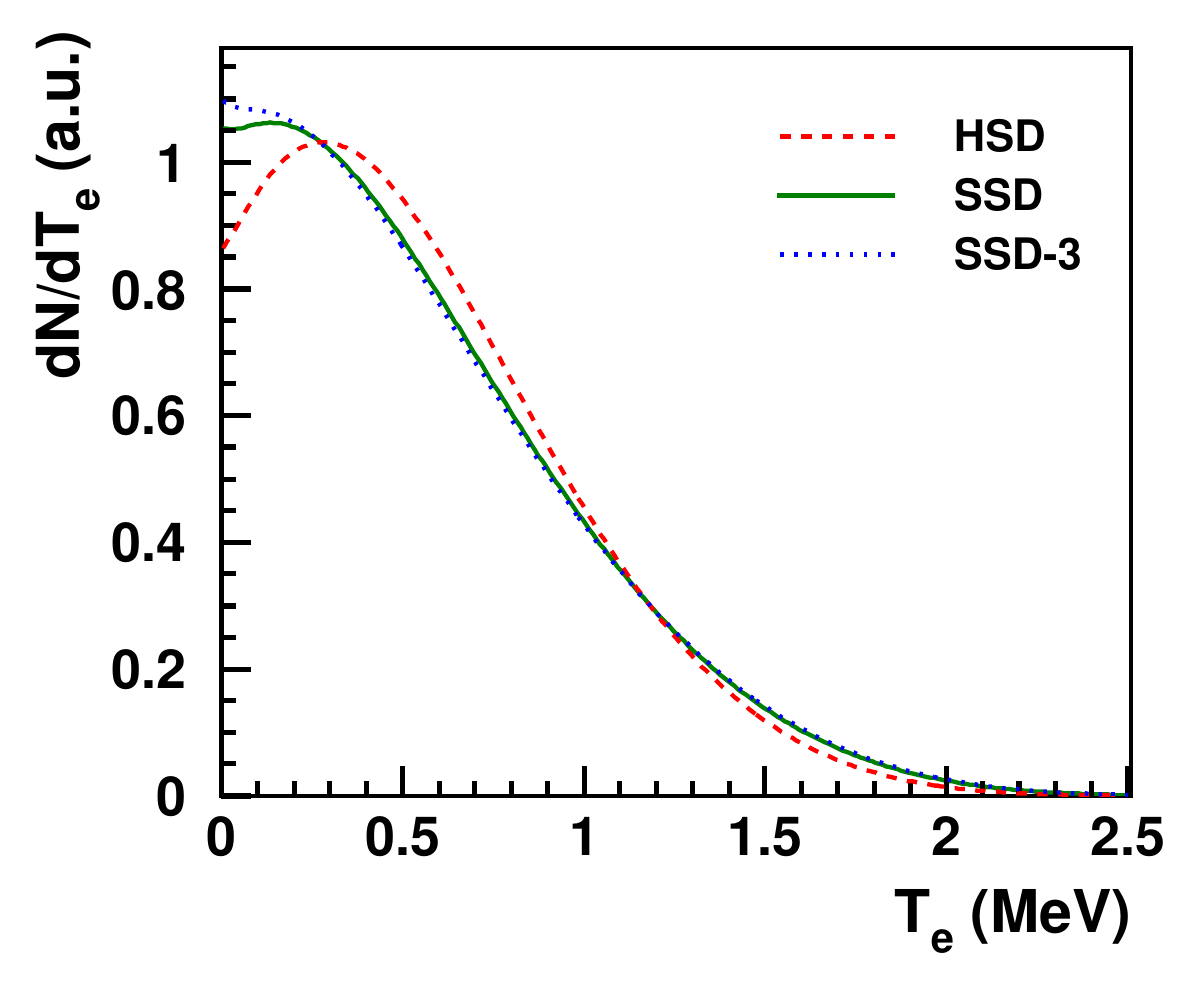}
\caption{Theoretical distributions of the individual electron 
kinetic energy for three models of $^{100}$Mo $2\nu\beta\beta$ decay: HSD, SSD and SSD-3.
}
\label{fig:ssd_theor}
\end{center}
\end{figure}

It was shown in~\cite{ssd_simkovic} that SSD and HSD models can be {\em directly} distinguished by making
high precision kinematics measurements of $2\nu\beta\beta$ decay products. 
The distribution of the individual electron energies 
was shown to have the most discriminating power, especially in the low energy part of the spectrum. 
Fig~\ref{fig:ssd_theor} shows the individual electron energy spectra for three nuclear models, 
with SSD-3 being a modification 
of the SSD model where a finer structure of intermediate states is accounted for~\cite{ssd3}. 

\begin{figure*}
\begin{center}
\includegraphics[width=0.45\textwidth]{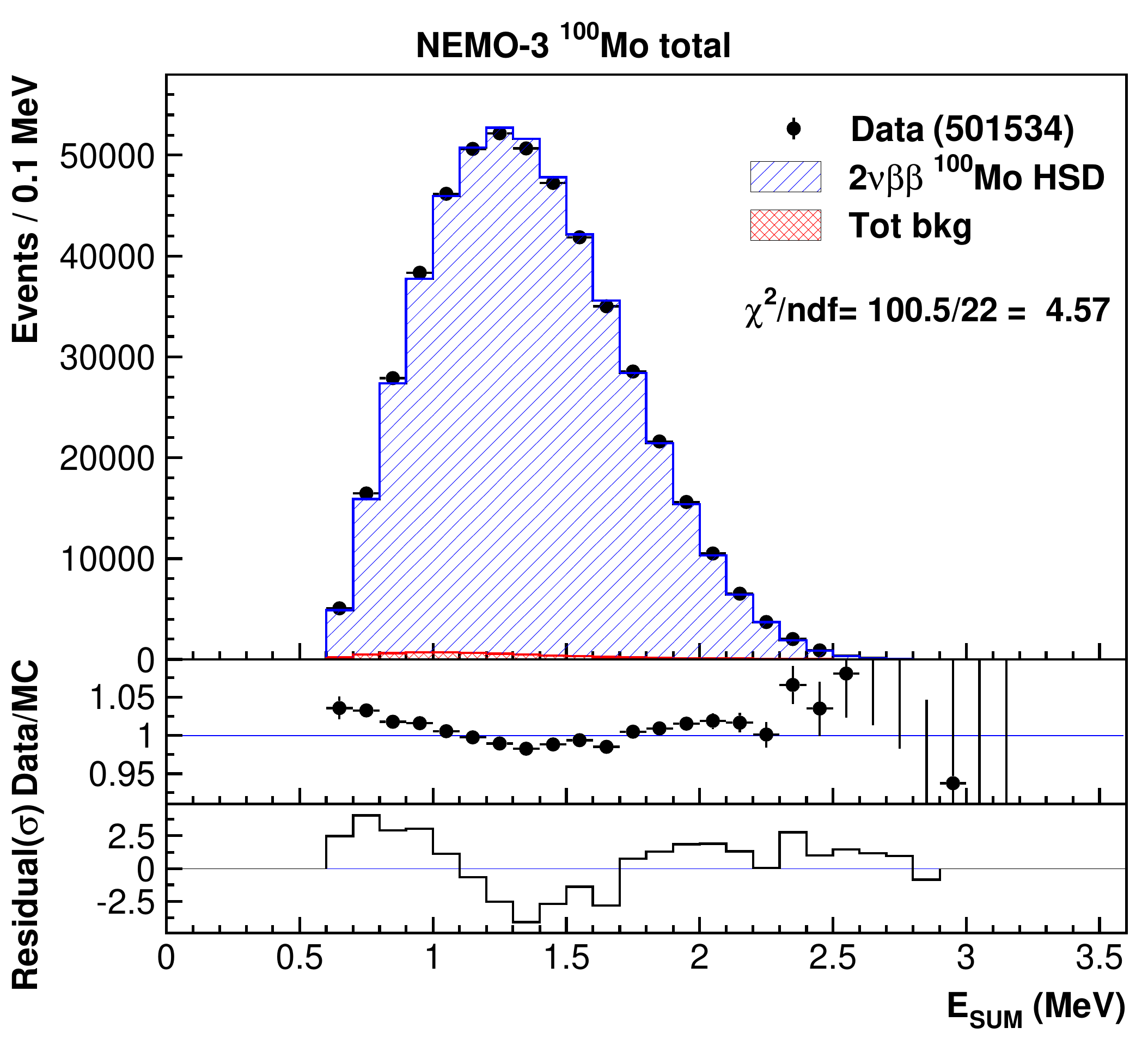}
\includegraphics[width=0.45\textwidth]{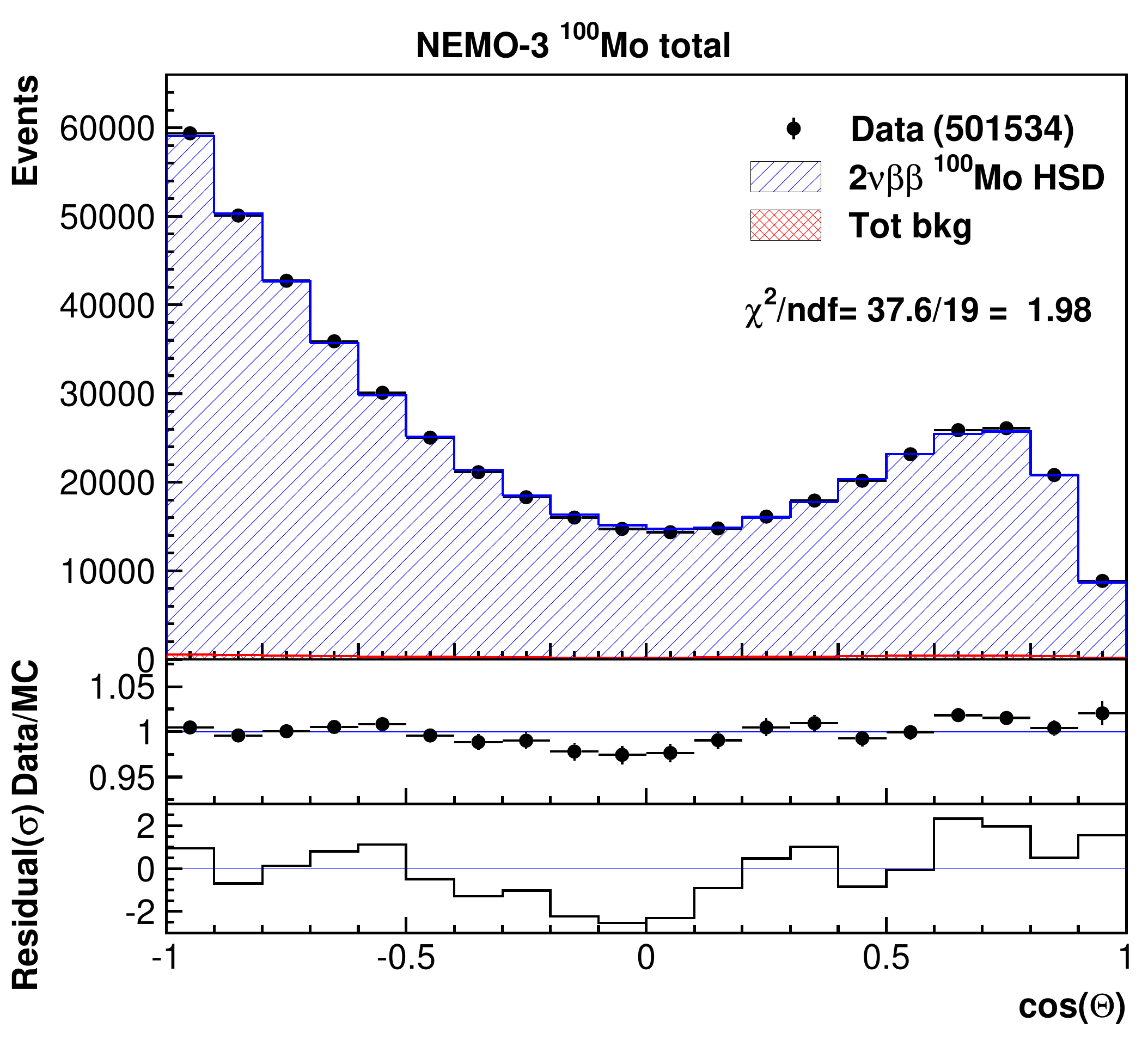}
\caption{Two-electron events. Energy sum and cosine of 
the angle between the two electrons for HSD model.
}
\label{fig:bb_HSD}
\end{center}
\end{figure*}

\begin{figure*}
\begin{center}
\includegraphics[width=0.31\textwidth]{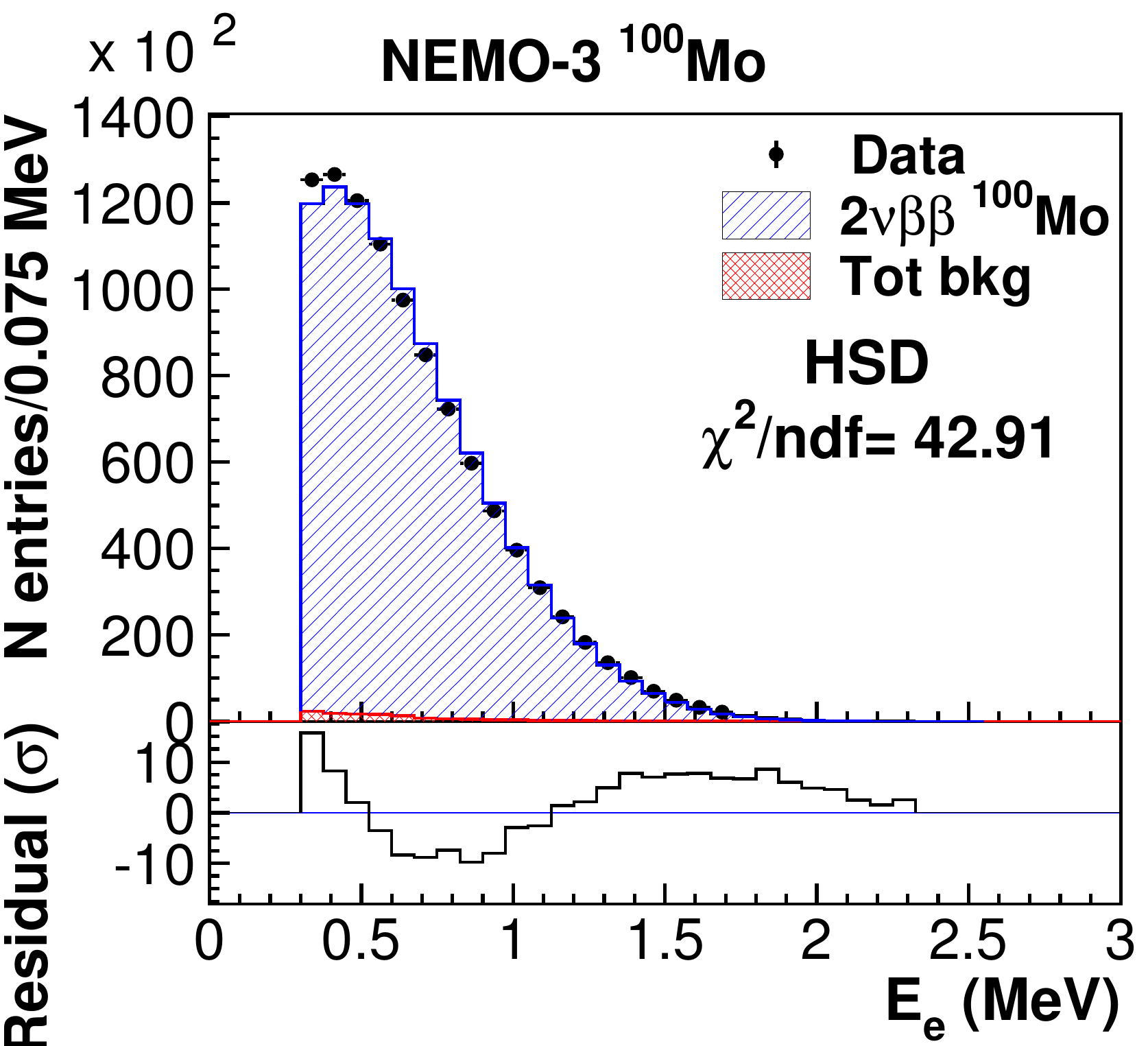}
~
\includegraphics[width=0.31\textwidth]{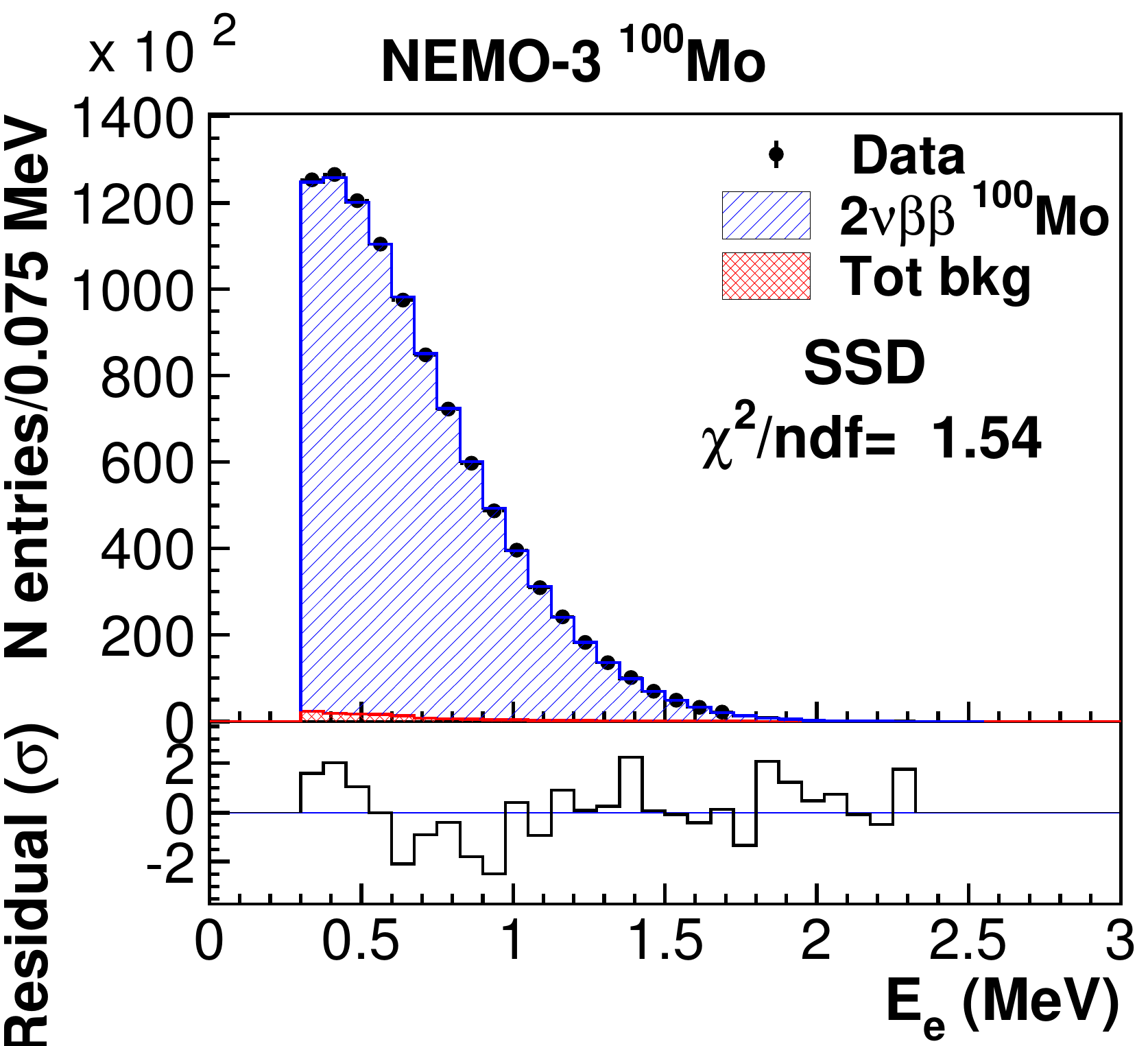}
~
\includegraphics[width=0.31\textwidth]{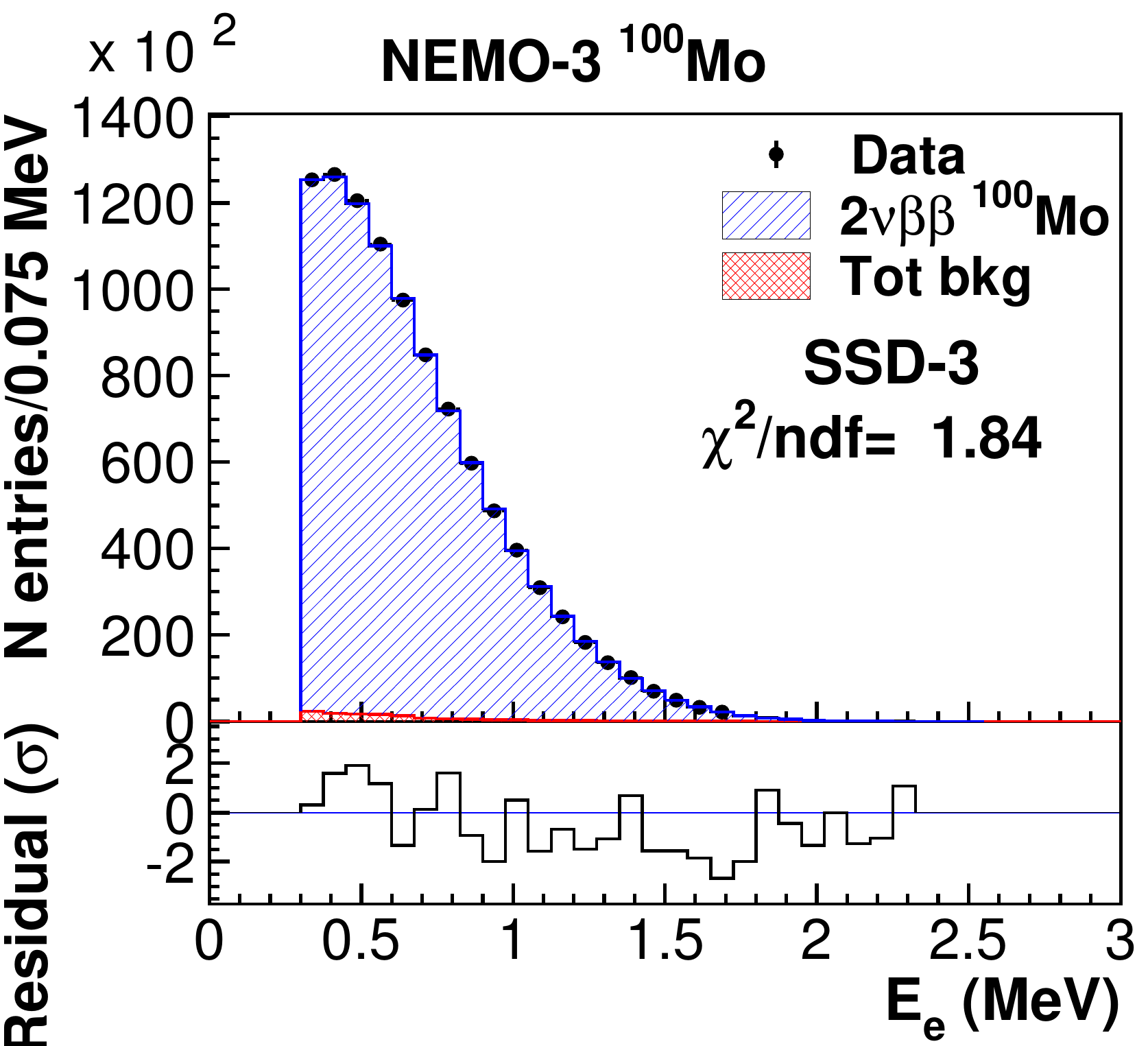}
\caption{Distribution of individual electron kinetic energy in the $\beta\beta$ channel from $^{100}$Mo foils compared with MC spectra under
the HSD, SSD and SSD-3 nuclear models. The HSD hypothesis is excluded ($\chi^{2}/\text{ndf} = 1159/27$) 
while the data are consistent with the SSD and SSD-3 models ($\chi^{2}/\text{ndf} = 41.5/27$ and $\chi^{2}/\text{ndf} = 49.7/27$ respectively).
}
\label{fig:bb_ssd}
\end{center}
\end{figure*}

Fig.~\ref{fig:bb_HSD} shows the energy sum and angular distribution of the final state electrons 
where the data are fitted with the HSD model. The tension between the data and the model is evident already 
from these distributions with $\chi^2$/ndf=4.57 ($p$-value=5.3$\cdot$10$^{-12}$) 
and $\chi^2$/ndf=1.98 ($p$-value=0.007) for the energy sum 
and angular distributions respectively. However, the strongest evidence comes from the single electron energy 
distributions shown in Fig.~\ref{fig:bb_ssd} for the three models, HSD, SSD and SSD-3, fitted to the data. 
It is clear from the distributions and $\chi^2$ values that the HSD model can be ruled out with high confidence 
while SSD and SSD-3 provide a fairly good description of the data. 

\begin{figure*}[p]
\begin{center}
\includegraphics[width=0.31\textwidth]{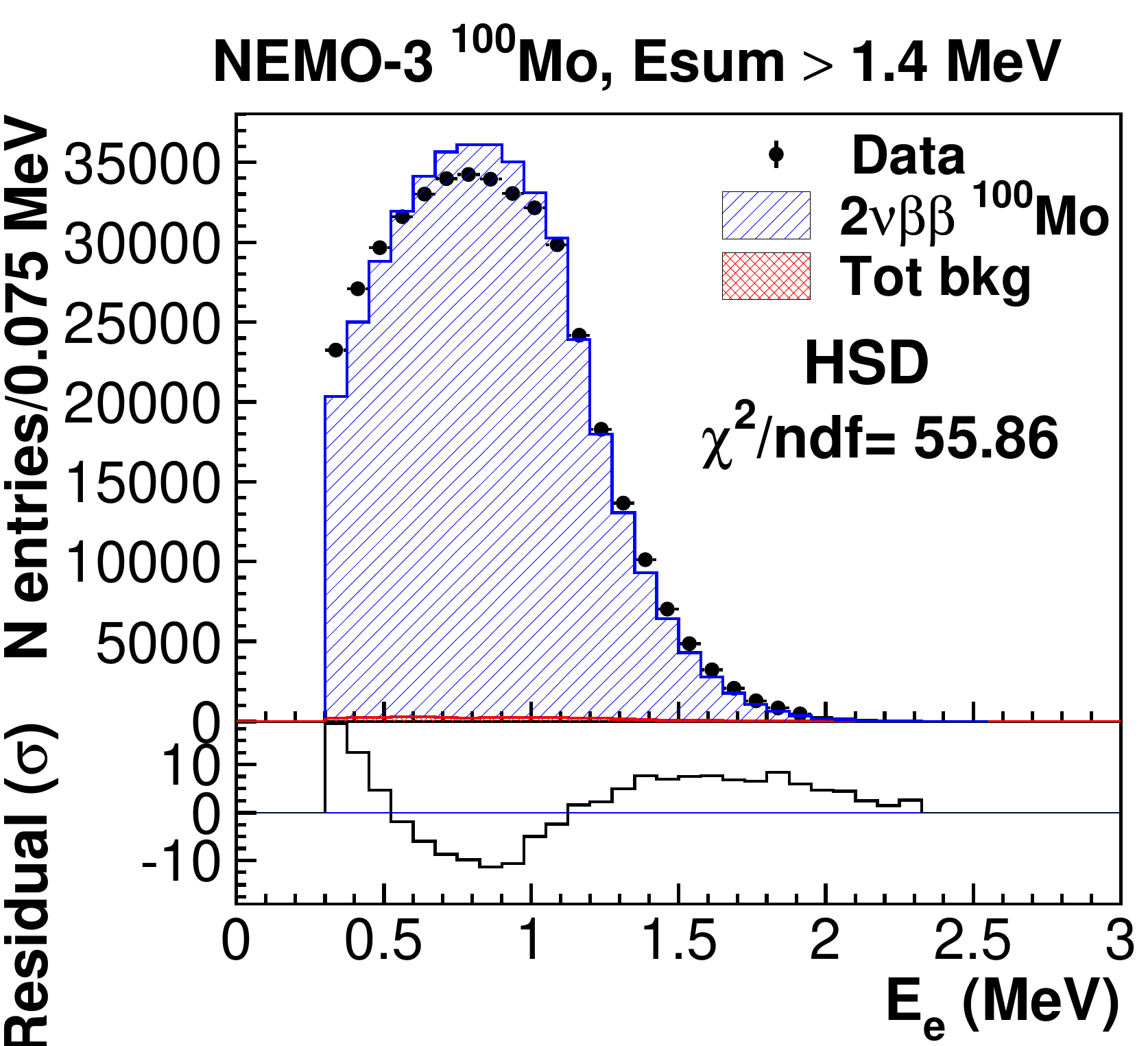}
~
\includegraphics[width=0.31\textwidth]{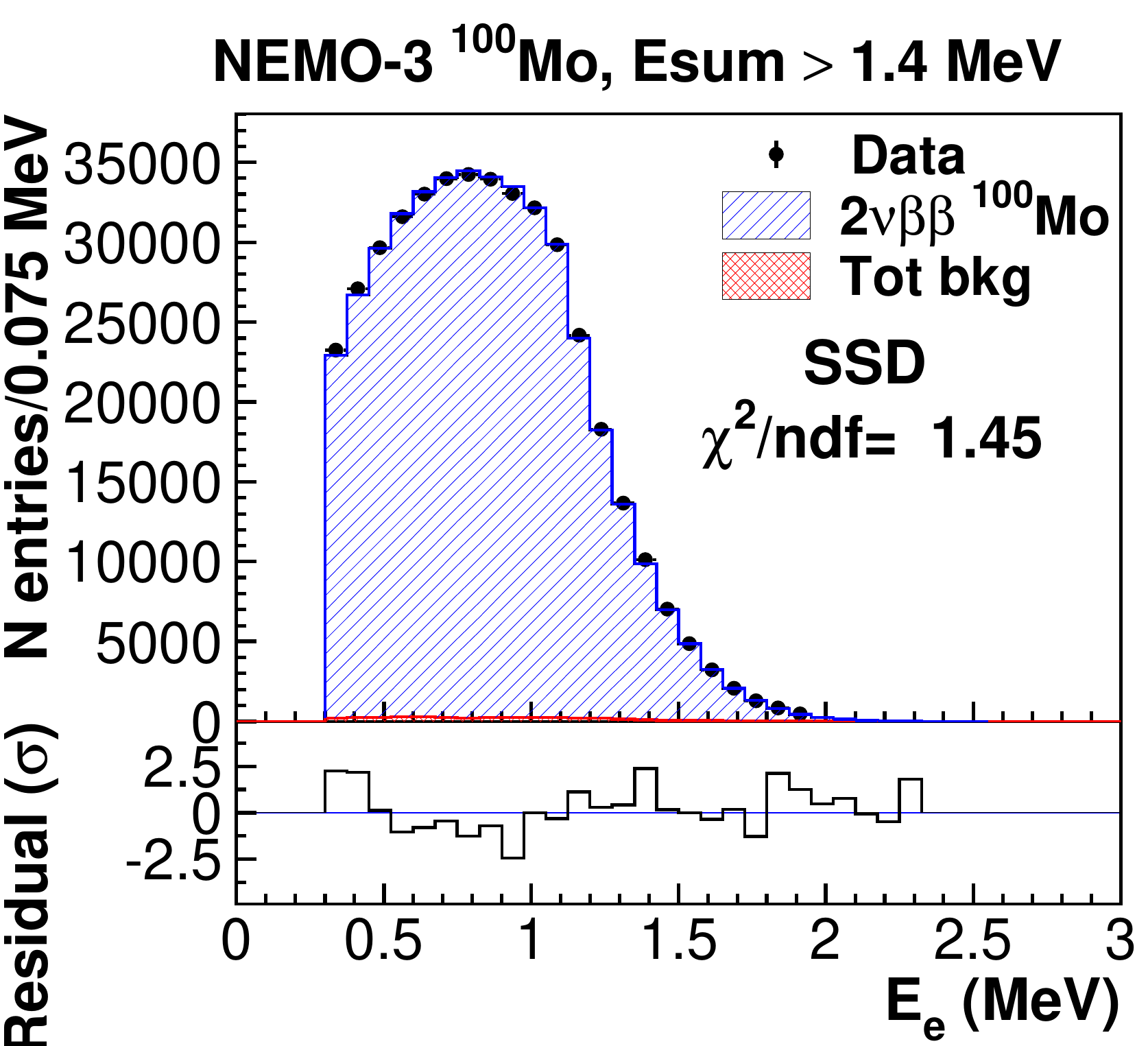}
~
\includegraphics[width=0.31\textwidth]{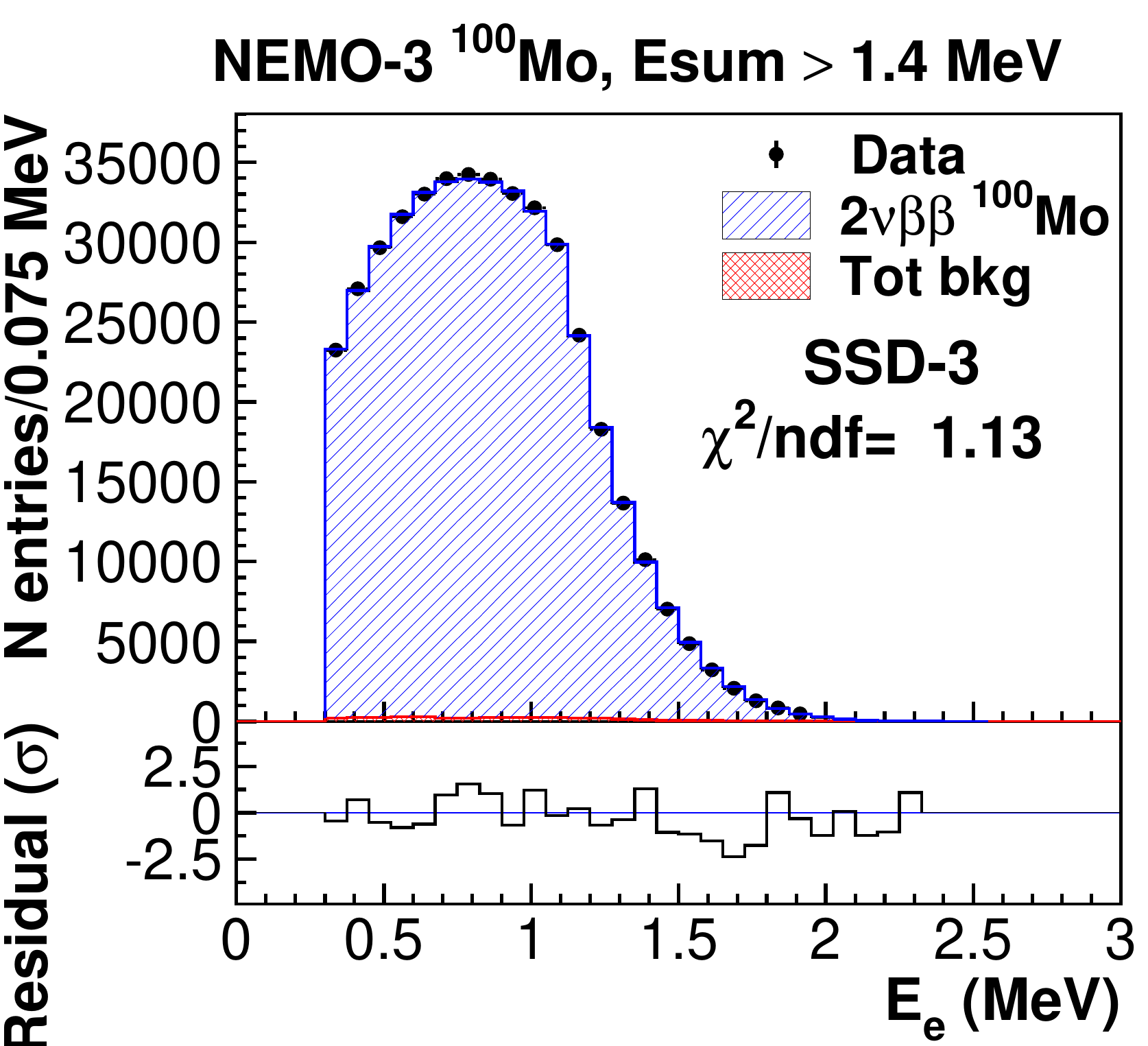}
\caption{Distribution of individual electron kinetic energy  in the $\beta\beta$ channel from $^{100}$Mo foils 
with the cut on the summed electron energy $E_{SUM} > 1.4$ MeV to maximise the signal-to-background ratio. 
The data are compared with MC spectra under
the HSD, SSD and SSD-3 nuclear models. The HSD hypothesis is excluded ($\chi^{2}/\text{ndf} = 1508/27$) 
while the data are consistent with the SSD and SSD-3 models ($\chi^{2}/\text{ndf} = 39/27$ and $\chi^{2}/\text{ndf} = 30.6/27$ respectively).
}
\label{fig:bb_ssd_etot14}
\end{center}
\end{figure*}

The difference between SSD and SSD-3 in describing the data is maximised 
with a cut on the electron energy sum of $E_{SUM} > 1.4$~MeV as shown in Fig.~\ref{fig:bb_ssd_etot14},
which also increases the signal-to-background ratio. There is a slight preference of 
the SSD-3 model over SSD in this case,
contrary to the results obtained without this cut 
demonstrated at Fig.~\ref{fig:bb_ssd}.
Due to systematic effects connected to the energy reconstruction and electron energy loss 
simulations discussed below
these two models cannot be discriminated against each other. The SSD is chosen as the baseline model 
and is used to estimate the $^{100}$Mo $2\nu\beta\beta$ half-life (see Section~\ref{sec:2vbb} and Fig.~\ref{fig:bb_energy}). 
We note that differences in the low energy part of the single electron spectra (Fig.~\ref{fig:ssd_theor}) 
affect the selection efficiency of $^{100}$Mo $2\nu\beta\beta$ events. Consequently, 
the measured half-life for the SSD model is 14\% shorter than the analogous result for the HSD model.
The SSD-3 model would give a 1.8\% shorter half-life than that of the SSD model.  


\subsection{Systematic uncertainties on $^{100}$Mo $2\nu\beta\beta$ half-life}
\label{sec:syst}

Apart from the statistical uncertainties on the fitted
number of signal events, 
the measurement of the $2\nu\beta\beta$ decay half-life is
subject to a number of systematic uncertainties.

The uncertainty on the reconstruction and selection efficiency including the detector acceptance effects is evaluated 
by carrying out dedicated calibrations with $^{207}$Bi sources whose activities were known with a 5\% uncertainty.
Consequently, the systematic error on the signal efficiency is taken to be 5\%. 

Limited presision of MC simulation program in modelling of multiple scattering processes 
and electron energy losses in 
molybdenum $\beta\beta$ source foils also contribute to the total systematic
error. Corresponding uncertainty is evaluated as the difference 
between the mean half-life value and the
values obtained with metallic (-2.3\%) and composite (+1.5\%) foils.

The 1.8\% half-life value difference between  the SSD and SSD-3 nuclear models is taken
as a systematic error due to the $^{100}$Mo $2\nu\beta\beta$ decay model.

The uncertainty on the energy scale translates into an error on the half-life measurement of 0.6\%. 

The $^{100}$Mo mass uncertainty gives directly the corresponding uncertainty 
of the half-life value and is estimated to be 0.2\%.

The error on the activities of external backgrounds, radon and the foil contamination 
with $^{214}$Bi and $^{208}$Tl is 10\% as shown in~\cite{0nu-PR}. The uncertainty 
on the backgrounds from $^{40}$K in the source foils as well as from $^{210}$Bi is estimated to be 4\%. 
The observed discrepancy in the $^{234m}$Pa decay scheme reported in~\cite{pa234m-new} and~\cite{pa234m-old} 
lead to a 30\% normalisation uncertainty on the activity from this isotope. 
The 7.5\% error on the rate of the $^{100}$Mo $2\nu\beta\beta$ decay to the excited states~\cite{Barabash-2nu} 
is also taken into account. Overall, due to a high signal-to-background ratio 
the uncertainty on all background contributions 
produces only a 0.2\% systematic uncertainty on the $^{100}$Mo $2\nu\beta\beta$ half-life determination.

The systematic uncertainties on the measured $2\nu\beta\beta$ $^{100}$Mo half-life
are summarised in Table~\ref{table:syst_err}. The individual sources of the systematic error 
are assumed to be uncorrelated and the total uncertainty is obtained to be  [$+5.6,-5.8$]\%. 
\begin{table}[hbt]
\caption{\label{table:syst_err}%
Summary of systematic uncertainties on the measured 
$2\nu\beta\beta$ $^{100}$Mo half-life}
  \begin{tabular*}{\columnwidth}{@{\extracolsep{\fill}}lc@{}}
    \hline\noalign{\smallskip}
 Source of uncertainty   &  Effect on $T_{1/2}^{2\nu}$ (\%) \\
\noalign{\smallskip}\hline\noalign{\smallskip}
 Absolute normalization of $\epsilon_{2e}$        & $\pm 5$  \\
 Thin source foil modelling         & [$+1.5,-2.3$]  \\
 $^{100}$Mo decay model               & $ \pm 1.8$  \\
 Energy calibration                 & $\pm 0.6$   \\
 $^{100}$Mo    mass                 & $\pm 0.2$ \\
 Background uncertainty             & $\pm 0.2$ \\
\noalign{\smallskip}\hline\noalign{\smallskip}
 Total                              & [$+5.6,-5.8$]\\
\noalign{\smallskip}\hline 
\end{tabular*}
\end{table}
The final value of the half-life for the $2\nu\beta\beta$ decay of $^{100}$Mo under the SSD model is:
\begin{equation}
T_{1/2} = \left[ 6.81 \pm 0.01\,\left(\mbox{stat}\right) ^{+0.38}_{-0.40}\,\left(\mbox{syst}\right) \right] \times10^{18}~\mbox{y}.
\end{equation}
This value is in good agreement with the world average value of $(7.1 \pm 0.4) \times 10^{18}$~y \cite{Barabash-2nu} 
and with a recent result obtained using low-temperature scintillating bolometers 
(Li$_{2}$$^{100}$MoO$_4$),
$[6.90 \pm 0.15(\mbox{stat}) \pm 0.37(\mbox{syst})]\times10^{18}$~y \cite{Armengaud-2017}.

\section{Search for new physics with continuous $^{100}$Mo $\beta\beta$ energy spectra}
Deviations in the shape of the $2\nu\beta\beta$ energy spectra can provide hints of new physics. 
Below we report on results of searches for physics beyond the Standard Model that can modify the 
two-electron energy sum distribution of the $^{100}$Mo $2\nu\beta\beta$ decay due to emission of Majoron bosons, 
the existence of a bosonic component in the neutrino states and possible Lorentz invariance violation. 

The shape of the two-electron energy sum distribution in various types of decays
is characterized by the spectral index $n$  ~\cite{Barnet}, being determined by
the phase space $G \sim (Q_{\beta\beta}-T)^n$, where $Q_{\beta\beta}$ is the 
the full energy released in the
decay minus two electron masses and T is the sum of 
kinetic energies of two emitted electrons.
The ordinary $2\nu\beta\beta$ decay has a spectral index of $n=5$. 
Any modification from this functional form can be an indication of new physics. 

A number of grand unification theories predict the existence of a massless or light boson which couples to the neutrino. 
Neutrinoless $\beta\beta$ decay can proceed with the emission of one or two Majoron bosons resulting in a continuous energy sum spectrum with spectral index $n \neq 5$. 
The decay accompanied by a single Majoron emission 
has $n=1,2$ and $3$, while models with two Majoron emissions predict $n=3$~and~$7$~ 
(see \cite{Maj-old} and references therein). 
The results for the neutrinoless $\beta\beta$ decay with the emission of a
Majoron corresponding to the spectral index $n=1$ have already been published in~\cite{0nu-short,0nu-PR}. 	
The Majoron-accompanied $0\nu\beta\beta$ decay modes with spectral indices $n=2,3$ and $7$ are considered here.

It was noted in~\cite{Bosonic-nu} that violation of the Pauli exclusion principle resulting in a bosonic component 
in the neutrino states can  be tested  by looking at the shape of the energy and 
angular distributions of the electrons 
emitted in $\beta\beta$ decay. For the two-electron energy sum distribution the corresponding index would be $n=6$.

Lorentz invariance is a fundamental symmetry. However, new physics at very high energies 
close to the Planck scale can manifest itself in small effects at low energies, 
including Lorentz invariance violation. 
Consequently, searches for non-Lorentz invariant effects have attracted active theoretical and experimental 
effort~\cite{LV1,LV2,LV3,LV4}.  
The possibility to test Lorentz invariance with $\beta\beta$ decay was discussed in ~\cite{LV5,LV6}. 
In case of $2\nu\beta\beta$ decay
the Lorentz invariance violation may be manifested as a 
modification of the conventional electron sum spectrum
due to an additional contribution of the Lorentz-violating perturbation
with a spectral shape of $n=4$.
\begin{figure}[htb]
\begin{center}
\includegraphics[width=0.49\textwidth]{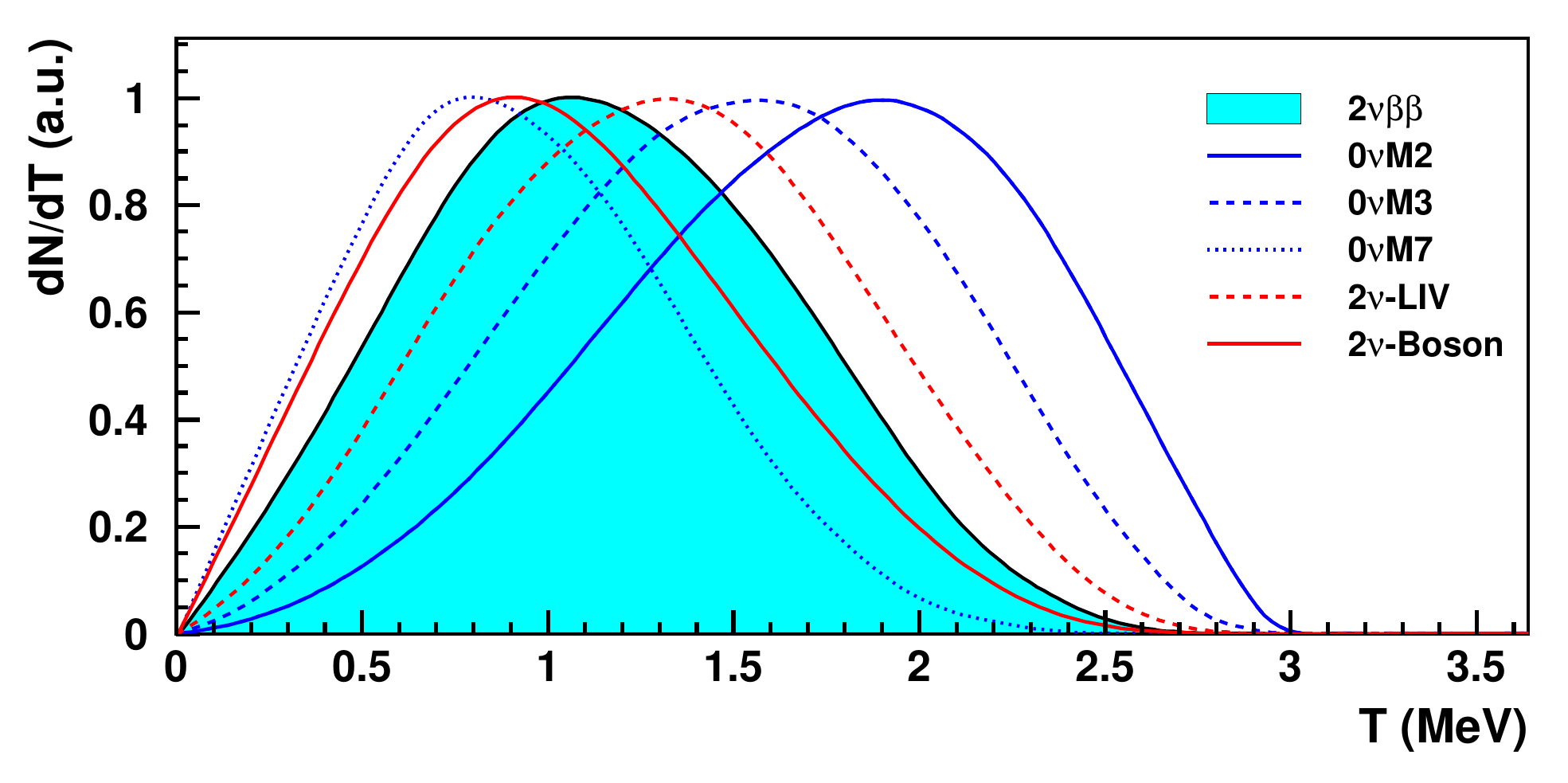}
\caption{Spectrum of the of kinetic energy sum of two electrons 
for the standard $^{100}$Mo $2\nu\beta\beta$ decay (spectral index $n=5$)
compared to the spectra for neutrinoless $\beta\beta$ decay with the emission 
of one or two Majorons $0\nu Mn$ ($n = 2, 3, 7$); shape of the perturbation to the standard $2\nu\beta\beta$ decay due to 
Lorentz invariance violation $2\nu$-$LIV$ ($n=4$) and 
spectrum for $2\nu\beta\beta$ decay with bosonic neutrino $2\nu$-Boson ($n=6$).
}
\label{fig:majoron}
\end{center}
\end{figure}

The theoretical distributions of the two-electron energy sum for different 
modes of $^{100}$Mo $\beta\beta$ decay
discussed above are shown in Fig~\ref{fig:majoron}. The difference in the 
shape of the distributions due to different 
spectral indices $n$ is used to evaluate possible contributions 
from physics beyond the Standard Model. 
No significant deviations from the expected $^{100}$Mo $2\nu\beta\beta$ 
spectral shape ($n=5$) have been observed 
and therefore limits on new physics parameters have been set using the full 
energy sum spectrum of the full $^{100}$Mo data set. 
The contributions of the $\beta\beta$ decay modes with spectral indices $n=2,3,6,7$ are constrained 
with a modified frequentist $CL_s$ method ~\cite{cls} using a profile likelihood fitting 
technique (COLLIE software package~\cite{collie}). 
A profile likelihood scan is used for the distribution with the spectral index $n=4$ 
in order to explore possibility of negative as well as positive Lorentz-violating perturbation.

The systematic uncertainties on background contributions discussed in Section~\ref{sec:syst}, 
the 5\% uncertainty on the detector acceptance and selection efficiency for signal, a possible distortion 
in the shape of the two-electron energy sum spectrum due to the energy calibration accuracy, as well 
as a 5\% error on the modelling of the energy loss of electrons 
are taken into account in limit setting without imposing 
a constraint on the normalization of standard $2\nu\beta\beta$ contribution.

The limits on the half-lives for different $0\nu\beta\beta$ modes with Majoron(s) emission, 
and for the bosonic neutrino admixture obtained with the $CL_s$ method are given in Table~\ref{table:CLfit2}.
\begin{table}[htb]
\caption{\label{table:CLfit2}%
Lower bounds on half-lives ($\times 10^{21}$~y) at $90\%$ C.L.
from $0\nu\beta\beta$ searches with Majoron emission (spectral indices $n=2,3,7$),
and searches for the bosonic neutrino admixture. 
The ranges in the expected half-life limits are from the $\pm1\sigma$ range of the systematic uncertainties 
on the background model, signal efficiency and distortions in the shape of the energy spectrum. 
}
  \begin{tabular*}{\columnwidth}{@{\extracolsep{\fill}}l|ccc|c@{}}
    \hline
                    & \multicolumn{3}{c|}{Expected}  &  \\
					  
Decay mode          &  $- 1\sigma$ & Median &  $+1\sigma$ &  Observed\\
\hline
Majoron n=2                   &$13$   & $9.2$ & $6.2$  & $9.9$ \\
Majoron n=3                   &$6.1$  & $4.3$ & $2.9$  & $4.4$ \\
Majoron n=7                   &$1.8$  & $1.3$ & $0.88$ & $1.2$ \\
$2\nu\beta\beta$ Bosonic $\nu$ &$1.7$  & $1.2$ & $0.83$ & $1.2$ \\
\hline    
\end{tabular*}
\end{table}
The half-life limits on the Majoron $0\nu\beta\beta$ modes are translated 
into the upper limits on the lepton number violating parameter $g_{ee}$, which is 
proportional to the coupling between the neutrino and the Majoron boson, using the relation,
\begin{equation}
1/T_{1/2} = |\langle g_{ee}\rangle|^m G |M|^2~,
\end{equation}
where $G$ is the phase space (which includes 
the axial-vector coupling constant $g_A$), 
$M$ is the nuclear matrix element, 
and $m=2(4)$ is the mode with the emission of one (two) Majoron particle(s).
The $M$ and $G$ values are taken from~\cite{Hirsch}. 
For the single Majoron emission and $n=3$, $M$ and $G$ are taken from~\cite{Barbero}.
There are no NME and phase space calculations available for $n=2$.

The upper limits on the Majoron-neutrino coupling constant $g_{ee}$ are shown
in Table~\ref{table:g_maj_com}. One can see that the NEMO-3 results presented here 
are the current best limits for $n=3$ and the single Majoron emission mode and are 
comparable with the world's best results from the EXO-200~\cite{exo-maj} 
and GERDA~\cite{gerda-maj} experiments for the other two modes. 

\begin{table}[htbp]
\caption{\label{table:g_maj_com}%
Upper limits on the Majoron-neutrino coupling constant $g_{ee}$ 
from NEMO-3 ($^{100}$Mo, this work) and EXO-200 ($^{136}$Xe)~\cite{exo-maj}
and GERDA ($^{76}$Ge)~\cite{gerda-maj} experiments. All limits are at 90\% C.L.
The ranges are due to uncertainties in NME calculations. 
}
  \begin{tabular*}{\columnwidth}{@{\extracolsep{\fill}}lcccc@{}}
    \hline\noalign{\smallskip}
 n & Mode & $^{100}$Mo  & $^{136}$Xe~\cite{exo-maj} & $^{76}$Ge~\cite{gerda-maj} \\
\noalign{\smallskip}\hline\noalign{\smallskip}
n=3   & $\chi^0$         &$0.013-0.035$      & $0.06$     &$0.047$\\
n=3   & $\chi^0\chi^0$   &$0.59 - 5.9$       & $0.6-5.5$  &$0.7-6.6$ \\
n=7   & $\chi^0\chi^0$   &$0.48 - 4.8$       & $0.4-4.7$  & $0.8-7.1$\\
    \noalign{\smallskip}\hline
\end{tabular*}
\end{table}

The contribution of bosonic neutrinos to the  $2\nu\beta\beta$-decay rate can be parametrised as~\cite{Bosonic-nu}: 
\begin{equation}
 W_{tot} = \cos^4 \chi W_f + \sin^4 \chi W_b,
\end{equation} 
 where $W_f$ and $W_b$ are the weights in the neutrino wave-function expression corresponding to the two fermionic 
and two bosonic antineutrino emission respectively. 
The purely fermionic, $T_{1/2}^{f}$, and purely bosonic, $T_{1/2}^{b}$, half-lives
are calculated under the SSD model to be~\cite{Bosonic-nu} :
\begin{equation}
T_{1/2}^f(0^+g.s. ) = 6.8\cdot 10^{18}~\mbox{y},~ ~
T_{1/2}^b(0^+g.s.) = 8.9\cdot 10^{19}~\mbox{y}.
\end{equation} 
Using the NEMO-3 half-life limit of $T_{1/2}^b(0^+g.s.) > 1.2\cdot 10^{21}$~y (Table~\ref{table:CLfit2}) 
an upper limit on the bosonic neutrino contribution to the $^{100}$Mo $2\nu\beta\beta$ decay 
to the ground state can be evaluated as:
\begin{equation}
\sin^2 \chi < 0.27 \,(90\%\,\mbox{C.L.}).
\end{equation} 
Although this limit is stronger than the bound obtained earlier in~\cite{Bosonic-nu},
the $2\nu\beta\beta$ transition of $^{100}$Mo to the ground state is not very sensitive 
to bosonic neutrino searches due to a small value of the expected bosonic-to-fermionic decay branching ratio
$r_0 (0^+g.s. ) = 0.076$. The $^{100}$Mo $2\nu\beta\beta$ decay to the first excited $2^+_1$ state 
has a branching ratio of $r_0 (2^+_1 ) = 7.1$~\cite{Bosonic-nu} and is therefore potentially more promising 
despite a lower overall decay rate. 
The current best experimental limit for this process is 
$T_{1/2}(2^+_1) > 2.5\cdot 10^{21}$~y~\cite{mo100_excited}.
This bound is still an order of magnitude lower than the theoretically expected half-life value of
$T_{1/2}^b(2^+_1) = 2.4\cdot 10^{22}$~y for purely bosonic neutrino, and 
two orders of magnitude lower than the corresponding expected value for purely fermionic neutrino, 
$T_{1/2}^f(2^+_1) = 1.7\cdot 10^{23}$~y~\cite{Bosonic-nu}.

The Standard Model Extension (SME) provides a general framework for Lorentz invariance violation (LIV)~\cite{LV1}.
In this model, the size of the Lorentz symmetry breakdown is controlled by SME coefficients that describe 
the coupling between standard model particles and background fields. Experimental limits have been 
set on hundreds of these SME coefficients from constraints in the matter, photon, neutrino and gravity sectors~\cite{LV1}.
The first search for LIV in $2\nu\beta\beta$ decay was carried out in~\cite{exo-200}. The two-electron energy sum spectrum 
of $^{136}$Xe was used to set a limit on the parameter $\mathring{a}^{(3)}_{of}$, which is related to a time-like component of this LIV
operator. The value of this parameter was constrained to be 
$-2.65\times 10^{-5}$~GeV~$ < \mathring{a}^{(3)}_{of} < 7.6\times 10^{-6}$~GeV
by looking at deviations from the predicted energy spectrum of $^{136}$Xe $2\nu\beta\beta$ decay ~\cite{exo-200}.

\begin{figure}[htb]
\begin{center}
\includegraphics[width=0.35\textwidth]{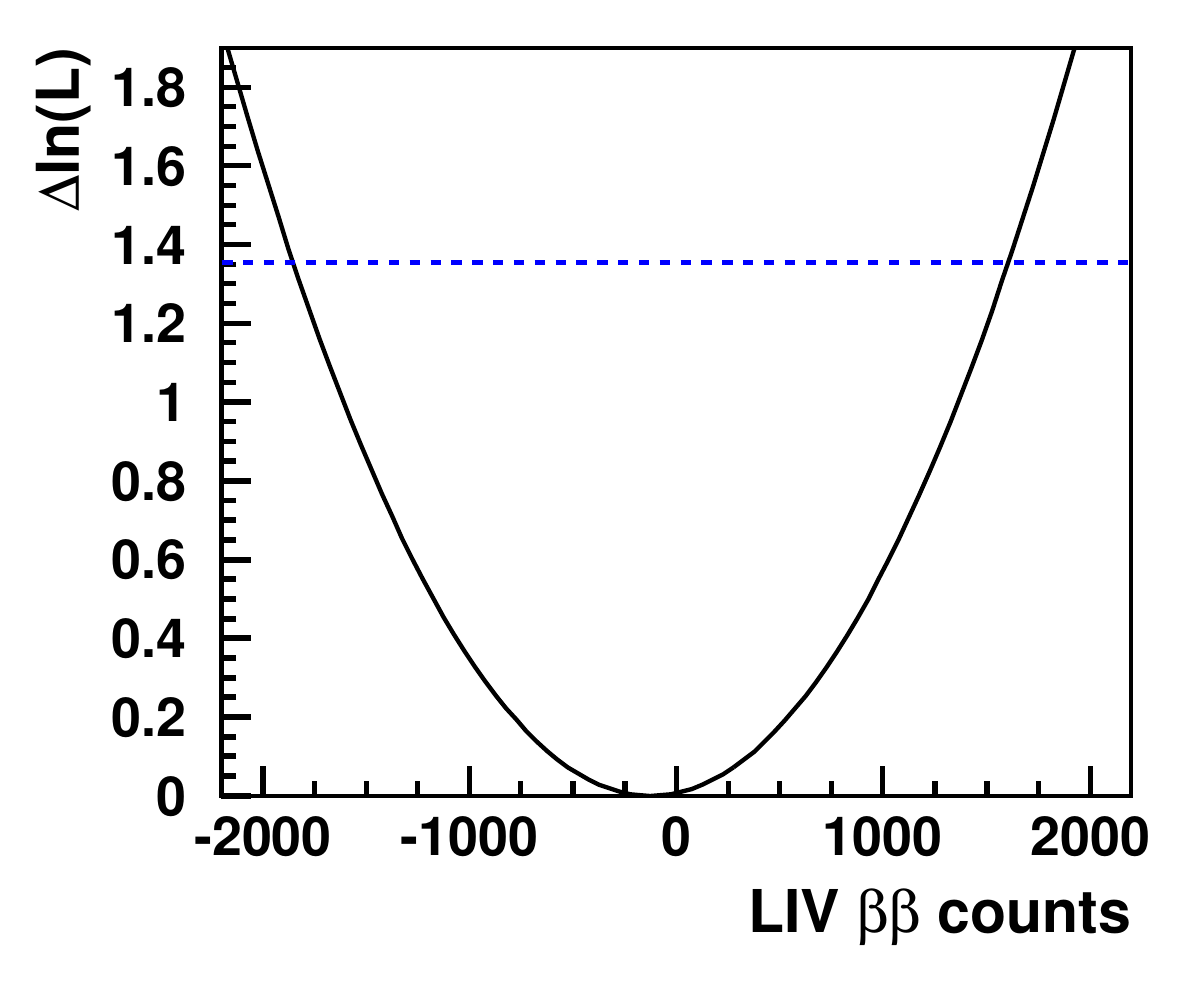}
\caption{Profile likelihood scan over observed  
two-electron LIV counts in  $^{100}$Mo $2\nu\beta\beta$ energy sum spectrum. 
The 90\% CL exclusion limit is shown with the dashed line. 
}
\label{fig:lorlim}
\end{center}
\end{figure}

In this work we adopt the same method, using the phase space calculations from ~\cite{Stoica}, 
and perform a profile likelihood scan over positive and negative contributions 
of LIV to two-electron events by altering the $^{100}$Mo $2\nu\beta\beta$ energy sum spectrum 
with positive and negative values of $\mathring{a}^{(3)}_{of}$. The result of this scan is shown in Fig.~\ref{fig:lorlim}.

The minimum of the profile log-likelihood function corresponds to $-135$ counts and is not 
statistically significant even at 1$\sigma$ level. The 90\% CL exclusion limit is shown in Fig.~\ref{fig:lorlim}
with the dashed line and gives $-1798$ and $1527$ events for negative and positive contributions  
to the deviation from the $^{100}$Mo $2\nu\beta\beta$ energy sum spectrum respectively. 
The corresponding constraint on $\mathring{a}^{(3)}_{of}$ is calculated using equations (2)-(6) in~\cite{exo-200}.
The result for  $^{100}$Mo obtained with a full set of NEMO-3 data is 
\begin{equation}
-4.2\times 10^{-7}~\mbox{GeV} < \mathring{a}^{(3)}_{of} < 3.5\times 10^{-7}~\mbox{GeV}\,(90\%\,\mbox{C.L.}).
\end{equation}
A summary of the best available constraints on LIV and CPT violation parameters can be found in 
compilation~\cite{LV1}.

\section{Summary}

The results of the $2\nu\beta\beta$ decay of $^{100}$Mo with the full data set of the NEMO-3 experiment
corresponding to a 34.3~kg$\cdot$y exposure are presented.
The summed energy of two electrons, the single electron energy and the angular distributions between the two electrons 
have been studied with an unprecedented statistical precision ($5\times10^5$ events).
The single electron energy distribution has been used to discriminate between different nuclear models
providing direct experimental input into NME calculations. The HSD model is excluded with high confidence,
while the SSD model is consistent with the NEMO-3 data. The corresponding half-life for 
the $2\nu\beta\beta$ decay of $^{100}$Mo is found to be 
\begin{equation}
T_{1/2} = \left[ 6.81 \pm 0.01\,\left(\mbox{stat}\right) ^{+0.38}_{-0.40}\,\left(\mbox{syst}\right) \right] \times10^{18}~\mbox{y}.
\end{equation}
Deviations from the expected shape of the $^{100}$Mo $2\nu\beta\beta$ energy sum spectrum 
have been studied to obtain constraints on parameters for physics beyond the Standard Model.
The most stringent upper limit to date has been obtained for the Majoron-neutrino coupling parameter 
$g_{ee}$ for the decay mode with a single Majoron particle emission and the spectral index $n=3$. 
For other $0\nu\beta\beta$ modes with two Majoron bosons emission a comparable sensitivity with the 
world's best limits has been achieved. 
The most stringent  constraints on the bosonic neutrino admixture and Lorentz
invariance violation in $2\nu\beta\beta$ decay have been set.

\section*{Acknowledgements}
We thank the staff of the Modane Underground Laboratory for their technical assistance in running the experiment. We are grateful to S.V.~Semenov for providing the spectra of SSD-3 model.
We acknowledge support by the grants agencies of the Czech Republic (grant number EF16\_013/0001733), 
CNRS/IN2P3 in France, RFBR in Russia (Project No.19-52-16002 NCNIL a), APVV in Slovakia (Project No. 15-0576), STFC in the UK and NSF in the USA.

\end{document}